\documentclass[journal]{vgtc}                     

\onlineid{1416}

\vgtccategory{Research}

\vgtcpapertype{theory/model}

\title{Manipulable Semantic Components:\\a Computational Representation of Data Visualization Scenes}
\author{%
  \authororcid{Zhicheng Liu}{0000-0002-1015-2759},
  \authororcid{Chen Chen}{0000-0003-3171-0657}, and 
  \authororcid{John Hooker}{0009-0009-9965-4015}
}

\authorfooter{
  \item
  	Zhicheng Liu, Chen Chen, and John Hooker are with University of Maryland, College Park.
  	E-mail: \{leozcliu, cchen24\}@umd.edu, jhooker@terpmail.umd.edu

}

\abstract{%
  Various data visualization applications such as reverse engineering and interactive authoring require a vocabulary that describes the structure of visualization scenes and the procedure to manipulate them. A few scene abstractions have been proposed, but they are restricted to specific applications for a limited set of visualization types. A unified and expressive model of data visualization scenes for different  applications has been missing. To fill this gap, we present Manipulable Semantic Components (MSC), a computational representation of data visualization scenes, to support applications in scene understanding and augmentation. MSC consists of two parts: a unified object model describing the structure of a visualization scene in terms of semantic components, and a set of operations to generate and modify the scene components. We demonstrate the benefits of MSC in three applications: visualization authoring, visualization deconstruction and reuse, and animation specification.
}


\graphicspath{{figs/}{figures/}{pictures/}{images/}{./}} 

\usepackage{mathptmx}                  

\usepackage{amsmath,amsfonts}
\usepackage{algorithmic}
\usepackage{algorithm}
\usepackage{array}
\usepackage{textcomp}
\usepackage{stfloats}
\usepackage{url}
\usepackage{verbatim}
\usepackage{graphicx}
\usepackage{cite}

\usepackage{mathptmx}                  
\usepackage{tabu}                      
\usepackage{booktabs}                  
\usepackage{xspace}
\usepackage{multirow} 
\usepackage{latexsym}
\usepackage[font=normalsize,labelfont=sf,textfont=sf]{subcaption}
\usepackage{pifont}
\usepackage{tabularx}
\usepackage{makecell}

\usepackage{upquote}
\usepackage{listings}
\usepackage{amssymb}
\usepackage{colortbl}
\usepackage{amsthm}
\usepackage{mdframed}

\definecolor{coolblack}{rgb}{0.0, 0.18, 0.39}
\definecolor{markcolor}{RGB}{191, 0, 0}
\definecolor{groupcolor}{RGB}{170, 121, 0}
\definecolor{layoutcolor}{RGB}{84, 130, 53}
\definecolor{enccolor}{RGB}{68, 114, 196}
\definecolor{constrcolor}{RGB}{216, 83, 214}
\definecolor{lightgray}{RGB}{212,212,212}
\definecolor{darkgreen}{RGB}{0, 153, 51}
\definecolor{eeeColor}{RGB}{238, 238, 238}

\lstdefinelanguage{JavaScript}{
  morecomment=[s]{/*}{*/},
  morecomment=[l]//,
  morestring=[b]",
  morestring=[b]'
}

\lstdefinestyle{js} {%
  morekeywords={typeof, new, catch, function, return, null, catch, switch, var, if, in, while, do, else, case, break, let},
  morekeywords={[2]{scene, import, mark, repeat, divide, encode, find, align, affix, legend, axis}},
  basicstyle={\linespread{0.95}\small\ttfamily},   
  columns=flexible,
  frame=tb,
  xleftmargin={0.45cm},
  numbers=left,
  stepnumber=1,
  firstnumber=1,
  numbersep=6pt,
  numberfirstline=true,
  numberstyle=\color{coolblack},
  identifierstyle=\color{black},
  keywordstyle=\color{darkjunglegreen}\bfseries,
  keywordstyle={[2]{\color{orange}\bfseries}},
  stringstyle=\color{darkjunglegreen}\ttfamily,
  commentstyle=\color{brown}\ttfamily,
  language=JavaScript,
  alsodigit={.:;},	
  tabsize=2,
  showtabs=false,
  showspaces=false,
  showstringspaces=false,
  extendedchars=true,
  breaklines=true,
  rulecolor=\color{gray},
  literate=%
  {Ö}{{\"O}}1
  {Ä}{{\"A}}1
  {Ü}{{\"U}}1
  {ß}{{\ss}}1
  {ü}{{\"u}}1
  {ä}{{\"a}}1
  {ö}{{\"o}}1
}

\newcommand{\eg}{{e.g.,}\xspace}
\newcommand{\ie}{{i.e.,}\xspace}

\newcommand{\densify}{{densify}\xspace}

\newcommand{\revision}[1]{\textcolor{black}{#1}}
\newcommand{\bpstart}[1]{\vspace{1mm} \noindent{\textbf{#1.}}}

\newcommand{\getRule}[4]{\vspace{0.35mm}
\begingroup
\color{coolblack}
\fontsize{8}{1\baselineskip}\selectfont 
$\mathsf{#1(#2):#3}$ $\leadsto$ $\mathsf{#4}$
\vspace{0.35mm}
\endgroup
}
  
\newcommand{\modificative}[1]{
\begingroup
\vspace{0.3mm}
\color{coolblack}
\fontsize{8}{1\baselineskip}\selectfont 
$\mathsf{#1}$
\vspace{0.15mm}
\endgroup
}

\newcommand{\ruleDef}[1]{
\begingroup
\vspace{0.3mm}
\color{coolblack}
\fontsize{8}{1\baselineskip}\selectfont 
$\mathsf{#1}$
\vspace{0.15mm}
\endgroup
}

\newcommand{\rc}[1]{$\mathsf{#1}$}

\newcommand{\cross}{\textcolor{red}{\ding{55}}}
\newcommand{\greenCheck}{\textcolor{darkgreen}{\ding{51}}}

\begin{document}



\maketitle


\newtheorem{definitioninner}{Definition}
\newenvironment{definition}[1][]
  {\begin{mdframed}[backgroundcolor=eeeColor,linewidth=0pt,innertopmargin=0pt]\begin{definitioninner}[#1]}
  {\end{definitioninner}\end{mdframed}}

\section{Introduction}
To create and render a data visualization, we must first assemble a scene---\revision{a data structure that includes all the visual elements and components, their attributes,  relationships, and constraints in the visualization.} 
A visualization scene can be manually assembled, for example, by using an XML-based markup language based on the SVG (Scalable Vector Graphics) specification \cite{noauthor_scalable_nodate}, or by drawing shapes on a canvas in a vector graphics editor. Since such processes are tedious for complex visualization designs, two popular approaches have been proposed to facilitate expressive scene assembly: scene manipulation libraries and declarative specifications. 
An example of a scene manipulation library is D3 \cite{bostock_d3_2011}, which uses the DOM (Document Object Model) to represent a visualization scene, and provides functions for programmers to select and modify scene elements.
Declarative specifications, on the other hand, decouple the description of desired visualization from scene assembly. Using languages such as ggplot2 \cite{wickham_ggplot2_2009} and Vega-Lite \cite{satyanarayan_vega-lite_2016}, programmers only need to specify their intents in declarative specifications, which are compiled into visualization scenes.

Both approaches have been successful in facilitating scene assembly. However, recent developments in various visualization applications postulate the need for going beyond scene assembly to support scene understanding and augmentation. For instance, visualization reverse-engineering involves an automated system deconstructing a scene to understand the semantic roles of visual objects and their relationships. Another example is interactive visualization authoring, where users can interactively edit \cite{bigelow_iterating_2017}, reuse \cite{cui_mixed-initiative_2022,chen_towards_2020}, and annotate \cite{ren_chartaccent_2017} static visualizations, as well as define interactive \cite{snyder_divi_2024} and animated behaviors \cite{ge_canis_2020,wang_animated_2021} through a graphical user interface.



Existing scene manipulation libraries and declarative specifications do not provide vocabularies at an appropriate level for such applications. Libraries like D3 use SVG as the scene graph model, which does not capture high-level semantic abstractions of visualization scenes 
\cite{chen_state_2023,ge_canis_2020}. Declarative specifications, by decoupling specification from scene assembly, hide details on how a scene is represented (\eg the hierarchical organization of the visual objects) and manipulated (\eg updating the number of marks based on the values of a data variable). They 
do not provide native support for directly referencing, selecting, and editing scene objects.  
Researchers thus resort to defining their own scene abstractions. For example, Chartreuse identified two primary mark structures (units and sequences) and six update operations (morph, move, fix,
recolor, repeat, and partition) to reverse-engineer and reuse infographics bar charts \cite{cui_mixed-initiative_2022}; Canis \cite{ge_canis_2020} defines a new scene model called dSVG to describe the data-enriched semantic components in SVG charts for animation authoring. Such efforts, however, are restricted to specific applications for a limited set of visualizations. A unified and expressive model of data visualization scenes for different  applications has been missing. 

To fill this gap, we present Manipulable Semantic Components (MSC), a computational representation of data visualization scenes. By computational representation, we refer to the format and methods that define how a data visualization scene is represented and manipulated within a computational system. MSC consists of two parts: a unified \textit{object model} describing the structure of a visualization scene in terms of semantic components, and a set of \textit{operations} to generate and modify the scene components. 

In the object model, we identify the following major types of semantic components: visual elements (\eg mark, glyph, collection), data scope, encodings (\ie mapping between data attribute and visual channel), algorithmic layouts (\eg grid, stack, packing), relational constraints (\eg alignment) and view configuration.  Each of these components has a set of associated operations. Specifically, we identify \textit{generative} operations that create or remove visual elements based on data (\eg repeat, divide, densify, classify, repopulate, and stratify), and \textit{modificative} operations that update visual properties of elements or relationships between elements (\eg apply visual encoding, customize scale, apply layout, apply constraint). 

We demonstrate the benefits of MSC and how it supports different applications in three case studies: visualization authoring,
visualization deconstruction and reuse, 
and animation authoring.
\section{Manipulable Semantic Components in a Diverging Stacked Bar Chart: An Example}\label{sec:example}

\begin{figure}[ht]
\centering
\subfloat[]{\includegraphics[width=\linewidth]{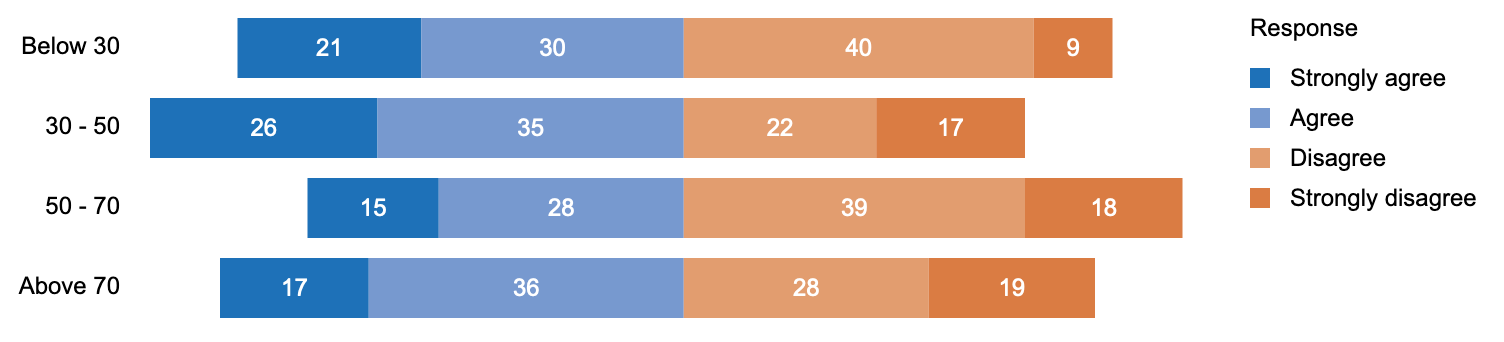} \label{fig:dsb_a}}
\hfil
\subfloat[]{\includegraphics[width=0.85\linewidth]{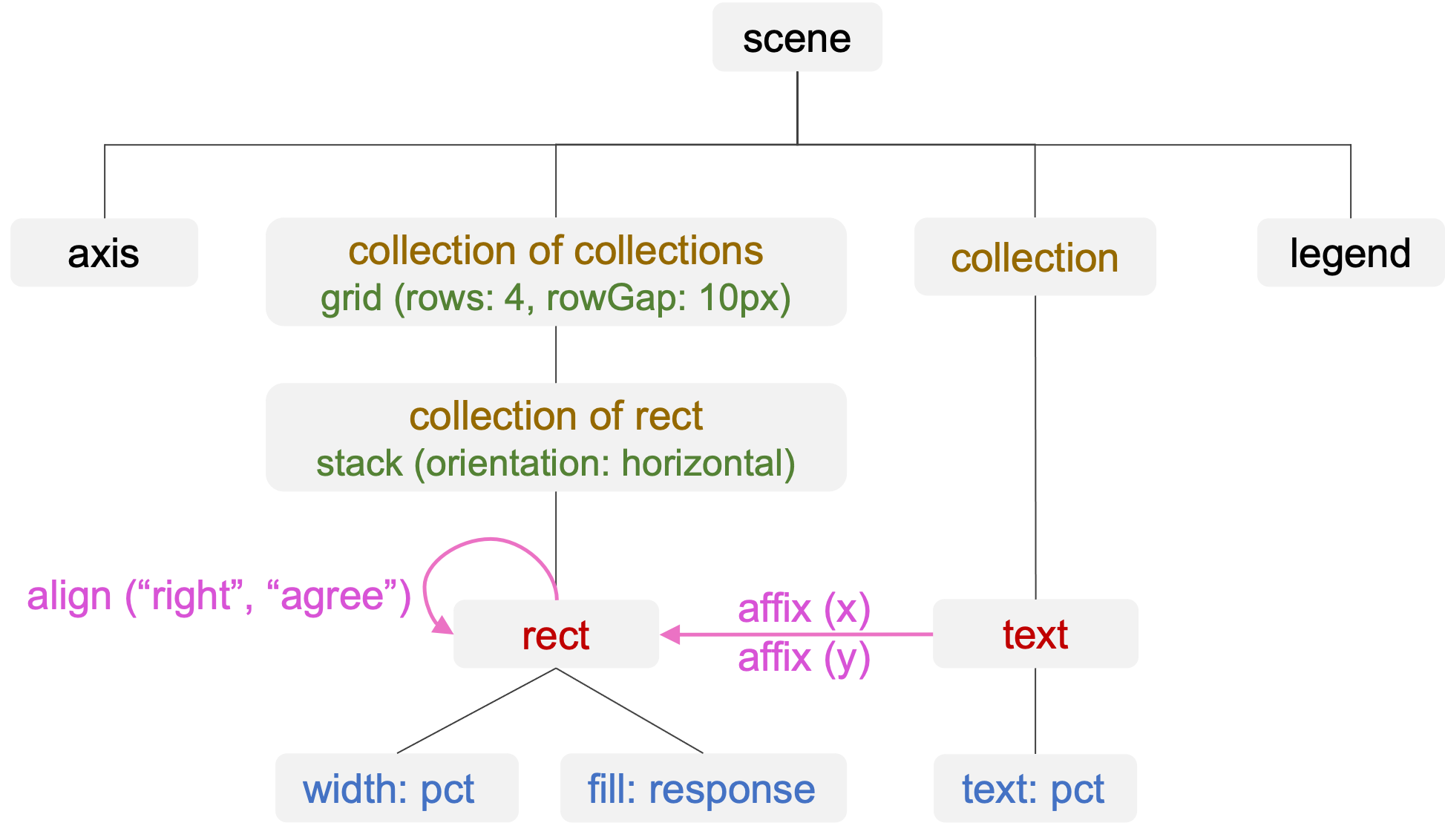} \label{fig:dsb_b}}
\caption{Manipulable semantic components in a diverging stacked bar chart. The scene abstraction not only captures hierarchical relationships between objects but also relationships like constraints and encodings.}
\vspace{-3mm}
\label{fig:dsb}
\end{figure}

For an example of manipulable semantic components, consider the diverging stacked bar chart in \cref{fig:dsb_a}. This chart visualizes a hypothetical dataset reporting people's opinions on a subject matter, broken down by age (below 30, 30 - 50, 50 - 70, above 70) and responses (strongly agree, agree, disagree, strongly disagree). 
The underlying data table consists of three columns: \rc{age}, \rc{response}, and \rc{pct} (percentage of people with a particular response within an age group). 
The scene of this chart consists of the following semantic components
(\cref{fig:dsb_b}):
{\textcolor{markcolor}{marks}} such as rectangles and texts are the graphical primitives; {\textcolor{groupcolor}{groups}} organize the rectangle marks by the \rc{age} values; {\textcolor{layoutcolor}{layouts}} specify the spatial relationships of rectangle marks within and across groups; \textcolor{enccolor}{encodings} bind data attributes (\eg \rc{pct}) to visual channels (\eg rectangle width); \textcolor{constrcolor}{constraints} enforce inter-object spatial relationships (\eg light blue rectangles are aligned to the right, text marks are affixed to the rectangle marks at the center). A component can have \textit{parameters}, e.g., a stack layout component has orientation and gap parameters. 

\begin{figure}[ht]
  \centering  \includegraphics[width=0.98\linewidth]{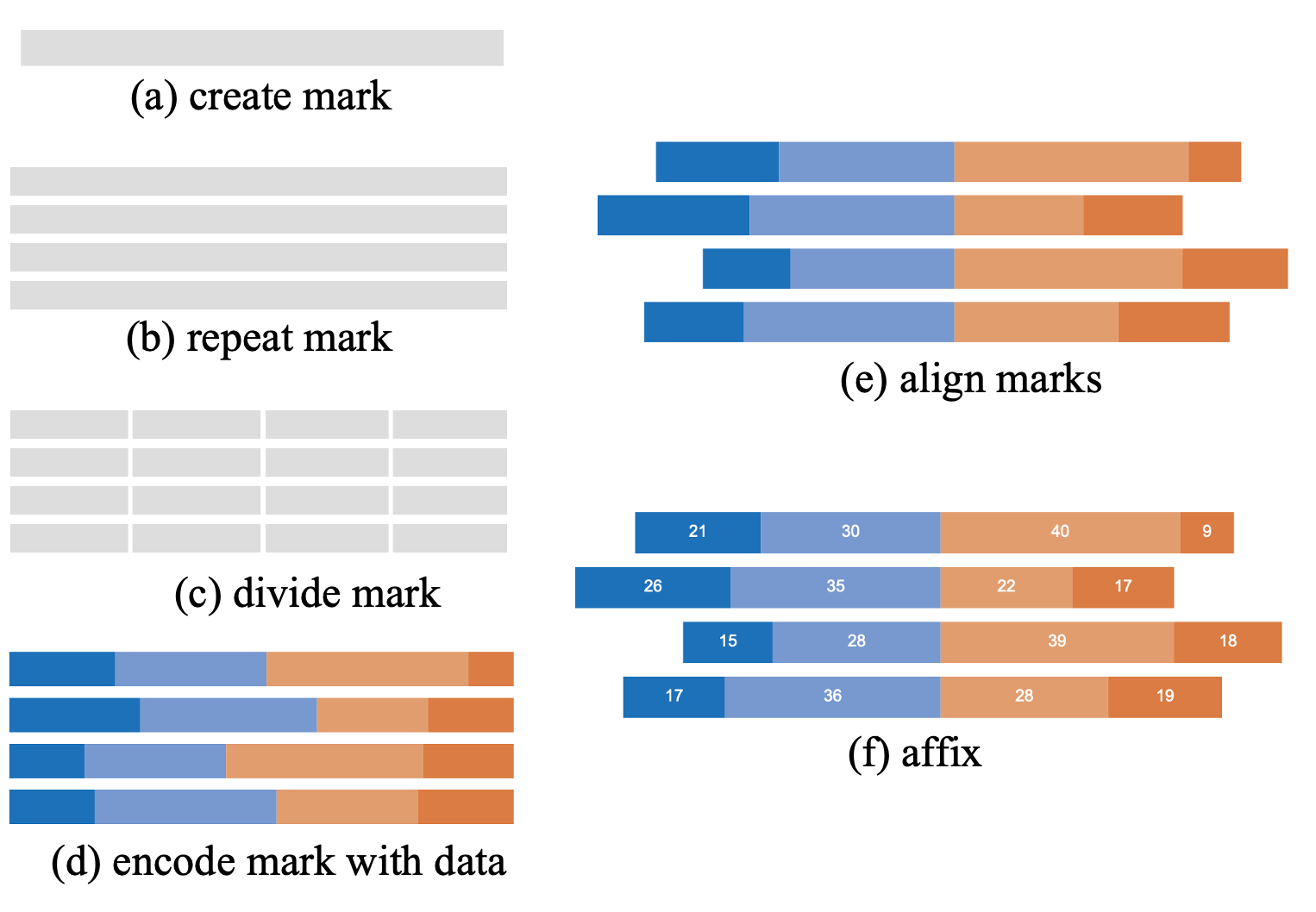} 
  \caption{Creating a diverging bar chart using the MSC operations. \label{fig:dsb_procedure}}
\end{figure}

In addition to describing the scene structure, MSC also describes the generative procedure to create and modify these components. \Cref{fig:dsb_procedure} shows major steps
of creating this chart using the MSC operations: 

\noindent(a)  \textit{create} a rectangle mark;  

\noindent(b) \revision{\textit{repeat} the rectangle by \rc{age}, resulting in a collection of four rectangles in a 4$\times$1 grid layout, each representing an age group}; 

\noindent(c) \revision{\textit{divide} each rectangle by \rc{response}, splitting it into four
rectangles in a horizontal stack layout, each representing a response type}; 

\noindent(d) 
\revision{
\textit{encode} the rectangles' width by \rc{pct} and fill color by \rc{response};} 

\noindent(e) \textit{align} the light blue rectangles to the right to better show the divergence of opinions; 

\noindent(f) \textit{repeat} an initial text item by \rc{pct} to create the labels, and \textit{affix} the position of the texts to the center of the rectangles. 

During this process, we can also freely \textit{resize} or \textit{translate} the objects. The encodings and constraints are enforced to ensure the design represents data faithfully. 
\vspace{-3mm}

\vspace{-3mm}

\section{Object Model: Semantic Components}\label{sec:components}
In MSC, a \textit{scene} consists of semantic components including visual elements, layouts, encodings, constraints, and view configuration. MSC focuses on two kinds of \textit{datasets}: tables and networks
\cite{munzner_visualization_2014}. A \textit{table} consists of a set of data \textit{items} (\ie rows), and each item contains values for a set of \textit{attributes} (\ie columns). A \textit{network} also contains items, in addition, it contains \textit{links} that connect the items. A tree is a special type of network and thus covered in MSC.  A dataset can be used across multiple scenes, and a scene can use multiple datasets. 

Given an initialized scene and a dataset, 
we can create or modify a visualization through the following semantic components: marks, data scope, groups, auxiliary visual elements, visual encodings and scales, algorithmic layouts, relational constraints, and view configuration. 


\subsection{Primary Visual Elements: Marks, Vertices \& Segments}

\begin{figure}[th]
  \centering
  \includegraphics[width=\linewidth]{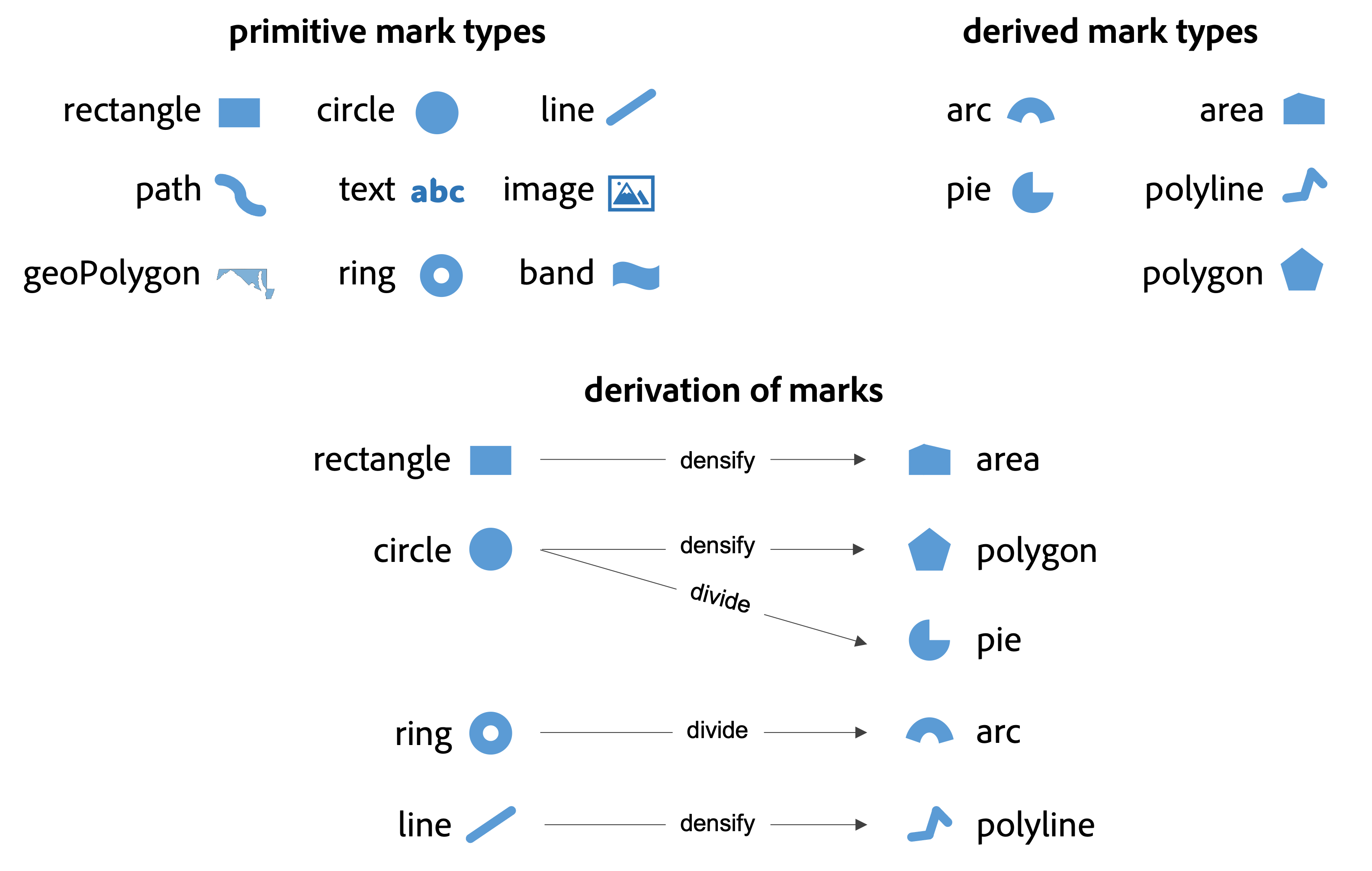} 
  \caption{Marks supported in in MSC and derivation of marks.\label{fig:marks}}
\end{figure}

Marks serve as the fundamental building blocks of data visualization. \cref{fig:marks} shows the fourteen mark types supported in MSC, as well as how certain types of marks can be derived by applying appropriate operations to some primitive marks (detailed in \cref{sec:procedure}). 
Marks can represent either items or links \cite{munzner_visualization_2014}. For instance, a line mark can either represent an item in a slope graph, or a link in a node-link diagram. 

\begin{definition}[Mark Type]
\revision{The type of a mark $m$ is defined as: \\
$type(m) \in \{\text{rectangle},~circle,~line,~path,~text,~image,~band,~area,$ \\
$~ring,~pie,~polyline,~arc,~polygon,~geoPolygon\}$}
\end{definition}

\vspace{-5mm}

Marks in the form of geometric shapes are composed of \textit{vertices} and \textit{segments}.
A rectangle, for example, consists of four vertices connected by four line segments; a polyline consists of multiple vertices connected by line or curve segments. 

\bpstart{Visual Channels} Each type of mark has a set of associated \textit{visual channels} (\eg width, height for rectangle, radius for circle) that determine the mark's visual appearance. Vertices and segments in a mark also have a set of associated visual channels (\eg x \& y positions of vertices in an area mark, y position of the top and bottom segments in a rectangle mark, and stroke width of segments in a polyline).


\subsection{Data Scope}
A mark represents one or more data items from a dataset. For example, each rectangle mark in \cref{fig:dsb_a} represents a distinct data item with a unique combination of \rc{age}, \rc{response}, and \rc{pct} values.  We define the data scope of a visual element $e$ to be the data item(s) represented by the element: $e.data$. 

\bpstart{Attribute Value and Aggregator} We use  $e.data[\alpha]$ to denote the value(s) of attribute $\alpha$ in element $e$'s data scope. The element can be a single mark or a group of marks. For instance, if $e$ refers to the bottom left blue bar in \revision{\cref{fig:dsb}a}, then 
$e.data[\text{"pct"}] = 17$; if $e$ refers to the group of four bars in the bottom row in \cref{fig:dsb}(f), then $e.data[\text{"pct"}] = [17, 36, 28, 29]$. 
In the latter case, since the attribute ($\text{"pct"}$) is quantitative, we can define an \textit{aggregator} (\eg \rc{max}, \rc{min}, \rc{count}, \rc{sum}) to compute a single value: $max(e.data[\text{"pct"}]) = 36$, and $min(e.data[\text{"pct"}]) = 17$. For simplicity, we use $e.data[\alpha]$ in the rest of this paper to refer to an element's scope attribute value, irrespective of the number of data items in the data scope.







\subsection{Primary Visual Elements: Groups}
MSC defines three types of groups: glyph, collection, and composite. Like marks, groups are visual elements with associated visual channels (commonly x and y positions) and can have data scopes.

\begin{figure}[th]
  \centering
  \includegraphics[width=\linewidth]{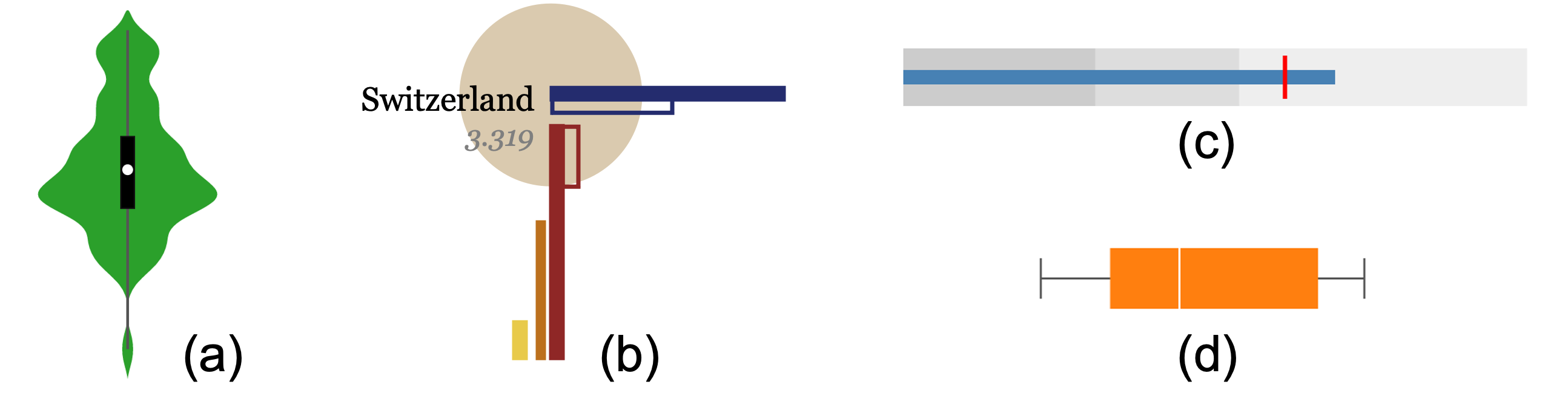} 
  \caption{Example glyphs in (a) a violin plot, (b) the ``Brain drain'' visualization \cite{accurat_brain_2014}, (c) a bullet chart, and (d) a box-and-whisker plot. \label{fig:glyph}}
\end{figure}

\bpstart{Glyph} Multiple marks can be grouped to create a glyph. \cref{fig:glyph} shows example glyphs in four different visualizations. 

\begin{definition}[Glyph]
A glyph has the following properties:

\noindent 1. All the marks in a glyph have the same data scope. 

\noindent 2. The data scope of a glyph is the same as the data scope of any of its member marks. 

\end{definition}
\vspace{-5mm}

\vspace{-5mm}

\begin{definition}[Glyph Type]
Given two glyphs \revision{$G_{i}$} and \revision{$G_{j}$}, they have the same type (i.e., type(\revision{$G_{i}$}) = type(\revision{$G_{j}$})) if the following  conditions are met: 

\noindent 1. $G_{i}$ and $G_{j}$ have the same number of constituent marks.

\noindent 2. for every mark $m_{a}$ in $G_{i}$, there exists a unique mark $m_{b}$ in $G_{j}$ where $type(m_{a}) = type(m_{b})$, and vice versa.
 
\end{definition}

\begin{figure}[th]
  \centering
  \includegraphics[width=\linewidth]{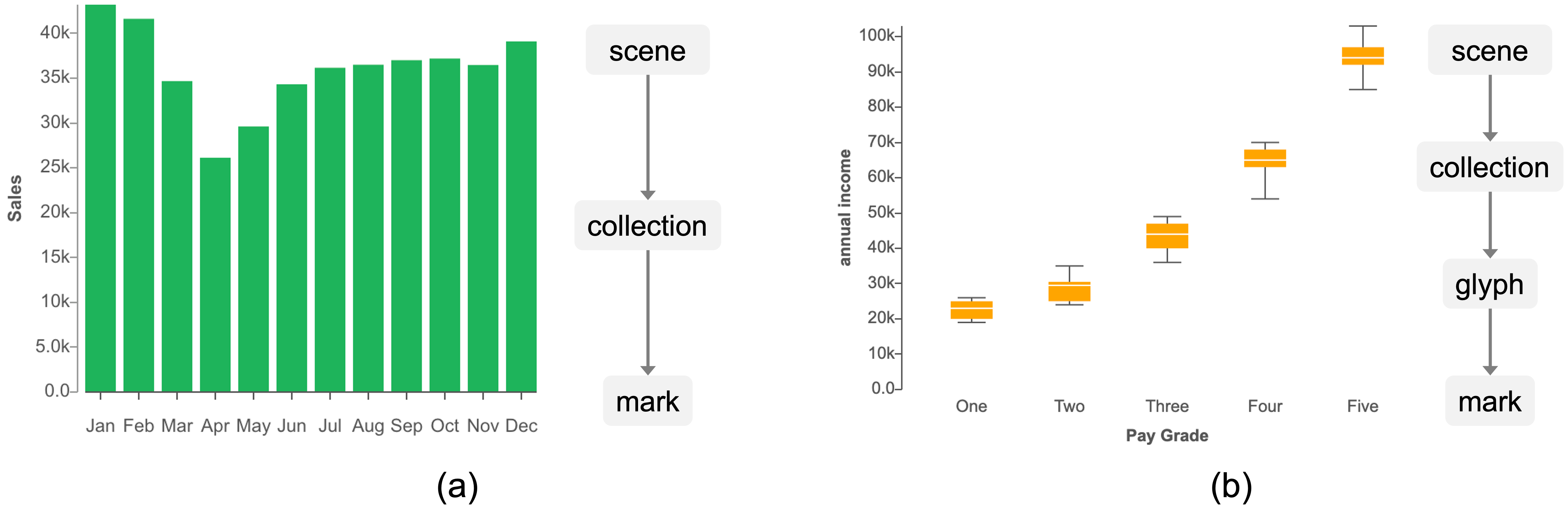} 
  \vspace{-5mm}
  \caption{(a) A collection of rectangles forms the main chart content of a bar chart; (b) a collection of glyphs forms that of a box plot.\label{fig:collection}}
  \vspace{-3mm}
\end{figure}

\bpstart{Collection} The main content of a data visualization is typically a collection of glyphs or marks (\cref{fig:collection}). No two members in a collection can share the same data scope. 
For example, each rectangle mark in  \cref{fig:collection}(a) represents a data item with a unique month value.
Collections can be nested: the members of a collection are collections. In \cref{fig:dsb}f, we have four collections of stacked bars, and they are the members of a larger collection that corresponds to the main chart content.

\begin{definition}[Collection]
A collection of visual elements has the following properties:

\noindent\revision{1. the members in a collection must be all marks, all glyphs, or all collections.}

\noindent\revision{2. any two members $e_{i}$ and $e_{j}$ in a collection must have the same type (per Definitions 1, 3, and 5): type($e_{i}$) = type($e_{j}$). 
} 


\noindent 3. the data scopes of all the members in the same collection share the same attributes.

\noindent 4. the data scopes of any two members $e_{i}$ and $e_{j}$ in a collection do not  overlap: $e_{i}.data~\cap~e_{j}.data = \varnothing$ . 

\noindent 5. the data scope of a collection $C$ is the union of the data scopes of its members: $C.data =~\bigcup_{e_{i} \in C}~e_{i}.data$ .
\label{def:collection}
\end{definition}

\vspace{-5mm}
\begin{definition}[Collection Type]
\revision{Two collections $C_{i}$ and $C_{j}$ have the same type (i.e., type($C_{i}$) = type($C_{j}$)) if any member $e_{i}$ from $C_{i}$ and any member $e_{j}$ from $C_{j}$ have the same type: type($e_{i}$) = type($e_{j}$).}
\end{definition}



\begin{figure}[th]
  \centering
  \includegraphics[width=\linewidth]{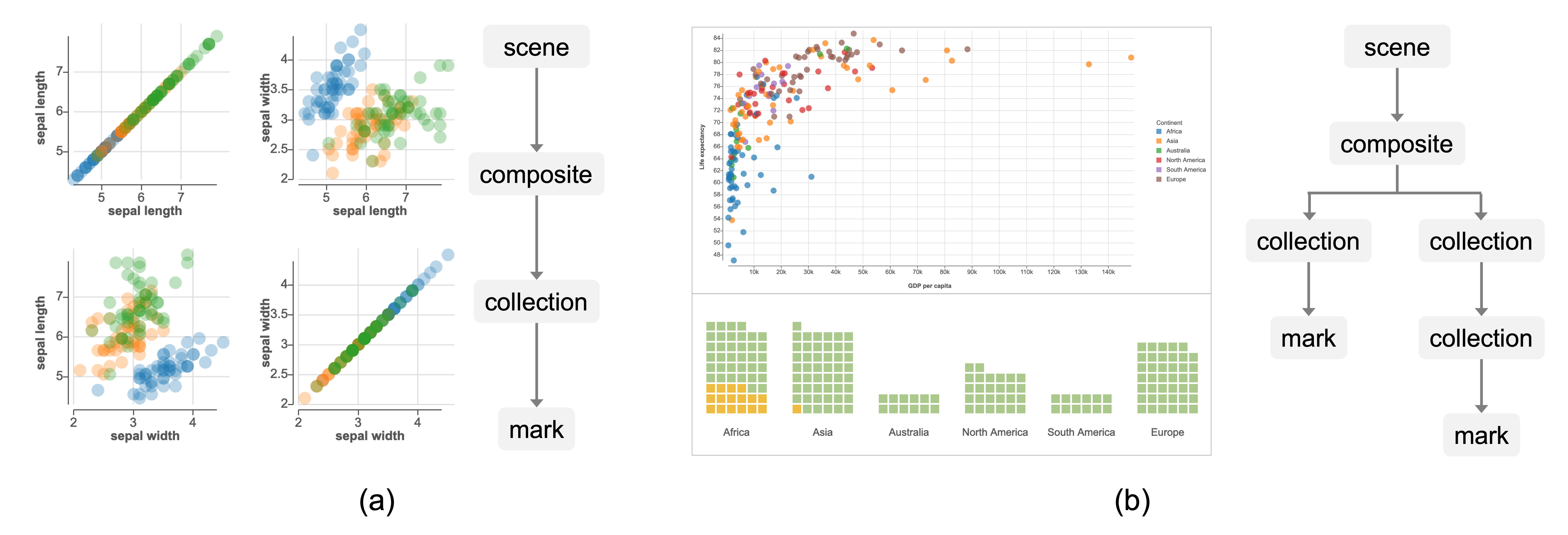} 
  \caption{(a) SPLOM is a composite of collections in the form of scatter plots, (b) a dashboard is a composite of multiple collections.\label{fig:composite}}
\end{figure}

\bpstart{Composite} A group of visual elements that is neither a glyph nor a collection is considered a composite.
\Cref{fig:composite} shows two example composites: (a) a scatter plot matrix (SPLOM), where each member (scatter plot) is a collection of circle marks, and (b) a dashboard consisting of two collections visualizing different fields of the same dataset: a scatter plot on top and small multiples of waffle charts. The waffle chart small multiples is also a nested collection. Note that the composite in either \cref{fig:composite}a or \cref{fig:composite}b, is not a nested collection because condition 3 in Definition~\ref{def:collection} is not satisfied. \revision{Members in a composite can be juxtaposed (\eg \cref{fig:composite}b) or superimposed (\eg  \cref{fig:scale}a, where a polyline and a collection of rectangles share the same x- and y-axes) \cite{javed_exploring_2012}.}

\revision{It is important to distinguish between glyph, collection, and composite  because (1) no generative operation in MSC (\cref{sec:join_operations}) applies to all types of groups (\eg we can classify a collection but not a glyph), and (2) knowing the relationships between different types of groups allows us to more precisely describe their creation and dissection (\eg when a composite consists of multiple collections, we can selectively manipulate a subset of the collections).}


\subsection{Auxiliary Visual Elements}
Auxiliary elements in a scene provide context and additional information for understanding the visualization. 

\bpstart{Reference Element} A visualization scene can contain reference elements like axes, legends, grid lines and labels, whose primary purpose is to facilitate the reading and interpretation of the visual appearances of marks, glyphs and collections. 


\bpstart{Annotation} Text, shapes, cues \cite{kong_internal_2017}, and images \cite{ren_chartaccent_2017} are often used as annotations to emphasize or explain specific data points, regions, or trends in a scene. Annotations provide additional context, offer clarity, and draw attention to specific parts of a visualization. \\

\subsection{Visual Encodings and Scales}
Visual encoding is a central component in data visualization, specifying the mapping of attribute values to visual channel values. In MSC, a visual encoding definition $enc$ consists of a visual element $e$, a visual channel $ch$, a data attribute $\alpha$, and a scale $\lambda$:

\ruleDef{enc~:=~\langle e, ch, \alpha, \lambda \rangle}

\begin{figure}[th]
  \centering
  \includegraphics[width=\linewidth]{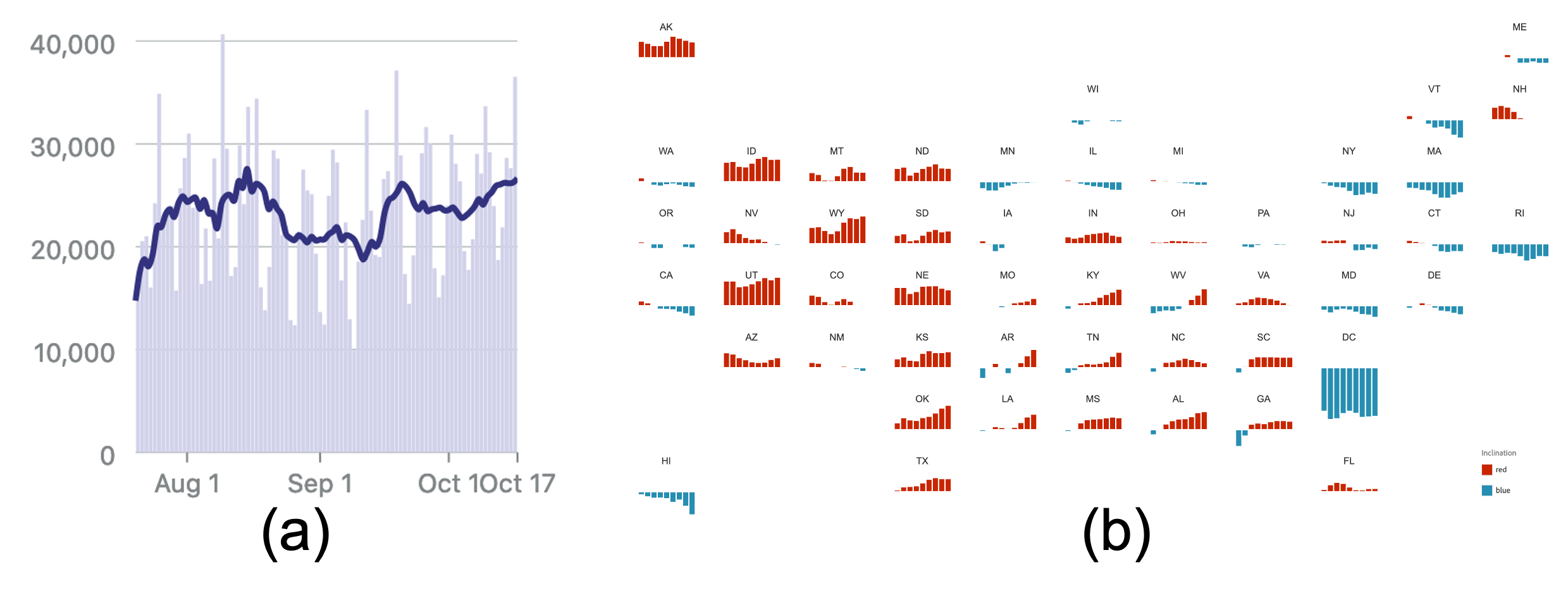} 
  \vspace{-5mm}
  \caption{(a) a line graph showing moving averages of COVID cases and a bar chart showing the raw case counts, (b) small multiples of bar charts showing how each state’s partisan lean has shifted over the years, where each bar chart is positioned based on its approximate geographic location of the corresponding state. 
  \label{fig:scale}}
\end{figure}

The visual element in the encoding definition can be a mark, a vertex or segment, or a group.
For example, in \cref{fig:dsb}a the width (visual channel) of each rectangle \textit{mark} (visual element) encodes \rc{pct} (attribute); in \cref{fig:collection}b, the y coordinate (visual channel) of each top whisker \textit{segment} (visual element) encodes the 75th percentile of annual income (attribute); in \cref{fig:scale}b, the x and y positions (visual channel) of each \textit{collection} (visual element) encode the approximate geographic locations (attribute) of each state.


For each visual encoding, MSC requires the definition of a \textit{scale}---a function specifying how the domain (data values) map to the range (visual channel values). For example, 
we have a scale in \cref{fig:dsb}a that maps \rc{pct} (attribute) values to the width (visual channel) of each rectangle mark: \ruleDef{\lambda(pct~value) = width~value}

Scales can be \textit{shared} or \textit{synced}. \Cref{fig:scale}a shows an example involving \textit{shared} scales: the visualization scene is a composite consisting of a polyline and a rectangle collection; the y-position encoding of the vertices in the polyline shares the same scale as the height encoding of the bars; the x-position encoding of the polyline vertices shares the same scale as the x-position encoding of the bars. Two encodings can share the same scale as long as the attributes have the same data type, sharing a scale does not require two encodings having the same attribute or channels. \Cref{fig:scale}b shows an example involving \textit{synced} scales. The visualization scene consists of 50 rectangle collections. Within each collection, the x-positions of the rectangles encode the \rc{year} attribute. The scales across the 50 collections have the same domain (values of the \rc{year} attribute), but the ranges expressed in terms of x coordinates are different. When editing this visualization, it is desirable that adjusting the scale range of one collection automatically updates the other collections to match. To achieve this, we can synchronize the scales across all 50 collections, so that updating the range extent of one scale will automatically adjust the range extents of the other scales.






\subsection{Algorithmic Layouts}
The spatial arrangement of visual elements is often achieved by visual encodings involving the position channels (\ie x and y coordinates). In addition to such encodings, algorithmic layouts are often used to specify the positions of visual elements. 
Examples of algorithmic layouts include but are not limited to: grid (\eg waffle chart), stack (\eg stacked bar chart), packing \cite{park_atom_2018} (\eg beeswarm plot), and spiral (\eg spiral plot). These layouts can be used to position not only marks and glyphs, but also collections (\eg a composite in the form of a dashboard can use a tiling layout to arrange its member collections; a nested collection such as a scatter plot matrix applies a grid layout to its member collections). Each of these algorithmic layouts has a set of associated \textit{parameters}: for example, a grid layout can specify the number of elements in a row/column and the direction of flow (\eg top to bottom or left to right). 

\vspace{2mm}
\subsection{Relational Constraints}

MSC includes relational constraints to specify inter-element relationships. The alignment constraint, for example, makes sure the anchors of elements are arranged in a straight line. In \cref{fig:dsb_a}, 
the light blue bars are aligned to the right, 
representing the divergence of opinions. To achieve this spatial arrangement, we can either align all the light blue bars to the right or all the light orange bars to the left.
Another type of constraint is the affixation constraint, which specifies the position of one element relative to an anchor element. In \cref{fig:dsb_a}, we can describe the positions of the white text labels as a result of affixing the text marks at the center of the rectangles. 

In addition, it is common to find constraints that specify the ordering of visual elements in a data visualization scene. For example, in a waffle chart, the ordering of rectangle marks in the collection together with the grid layout parameters determines the eventual visual appearance; in a connected scatter plot, the order of vertices on a polyline is determined by a temporal attribute and determines how the vertices are connected by segments. Ordering can also be specified along the z-dimension, which determines the order in which overlapping visual elements are drawn.  

%


\subsection{View Configuration}
The view configuration component provides a vantage point of a data visualization scene with configurable attributes like
focus point (center of the view port), field of view (extent visible through the
view port), zoom level, and rotation. Visualizations such as MatrixWave \cite{zhao_matrixwave_2015} and GeneaQuilts \cite{bezerianos_geneaquilts_2010}, for example, apply rotation to the scenes. 

\subsection{Alternative Scene Descriptions}
Three of the semantic components described above can determine the spatial positions of visual elements: visual encodings, algorithmic layouts, and relational constraints. It is possible to describe the same data visualization scene in multiple ways, each involving a different set of components. For example, we can describe the vertical bar chart in \cref{fig:collection}a as a collection of rectangles with a grid layout (one row, 12 columns) where the rectangles are sorted by the \rc{month} attribute, or a collection of rectangles whose x positions are bound to the \rc{month} attribute. Both descriptions are valid, but we should not include both the grid layout and the x encoding in the same description to describe the spatial arrangement of the bars along the horizontal direction.







%



\section{Procedure: Generative \& Modificative Operations}

\label{sec:procedure}
Given the definitions of semantic components in \cref{sec:components}, MSC \revision{includes} two kinds of operations on these components: 1) \textit{generative} operations that create new visual elements (\ie marks, glyphs, collections, composites) or remove existing ones based on data items and/or links, and 2) \textit{modificative} operations that modify existing visual elements' properties or change the relationships between them. 

\cref{sec:initialization_operations} and \cref{sec:join_operations} discuss generative operations: each operation consists of a descriptive name, an input visual element, an output element, and parameters. \revision{Most generative operations in MSC require all these parts to be specified. The input element and parameters may be optional for a few operations in \cref{sec:initialization_operations}}. We use the following form to represent a generative operation, similar notations have been used by generative grammars in linguistics \cite{chomsky_syntactic_2002} and shape modeling \cite{muller_procedural_2006}. 

\getRule{name}{parameters}{input~element}{output~element}

\vspace{0.8mm}

\cref{sec:channel_operations}, \cref{sec:spatial_operations}, and \cref{sec:camera_operations} discuss modificative operations,
where each operation \revision{is expressed either as \textit{value assignment} in the form of \modificative{element.property = value}, where a visual element's property is set to a constant, or as a \textit{mathematical constraint} in the form of \modificative{element_{1}.property = f(element_{2}.property, element_{3}.property, ...)}, where a function $f$ defines the relationships between the property values of multiple elements, usually expressed using arithmetic operators.
}







\subsection{Generative Operations: Initialize Data and Elements}
\label{sec:initialization_operations}








The first steps of creating a data visualization involve initializing a scene and importing data. In the operations below, 
\textcolor{coolblack}{$\mathsf{:=}$} denotes ``defined as'', \textcolor{coolblack}{$\mathsf{+}$} denotes ``one or more'', and \textcolor{coolblack}{$\mathsf{|}$} denotes ``or'':

\getRule{createScene}{}{}{scn}

\getRule{importData}{csvFile}{}{table := items~and~attributes} 

\getRule{importData}{jsonFile}{}{network := items,~attributes~and~links} 


\noindent We can then create primitive marks by specifying their types and properties, a scene is required as the input object: 

\getRule{createMark}{type,props}{scn}{mark:= rect|circle|line|path|txt|...} 

\noindent One or more marks can be grouped to create a glyph:

\getRule{createGlyph}{}{mark+}{glyph}

\subsection{Generative Operations: Join Visual Elements with Data} 
\label{sec:join_operations}
With primitive marks and data tables, we can now derive additional types of marks and create collections through graphics-data join operations. MSC specifies six such operations: repeat, divide, densify, classify, repopulate, and stratify. 


 \begin{figure}[th]
  \centering
  \includegraphics[width=1.0\linewidth]{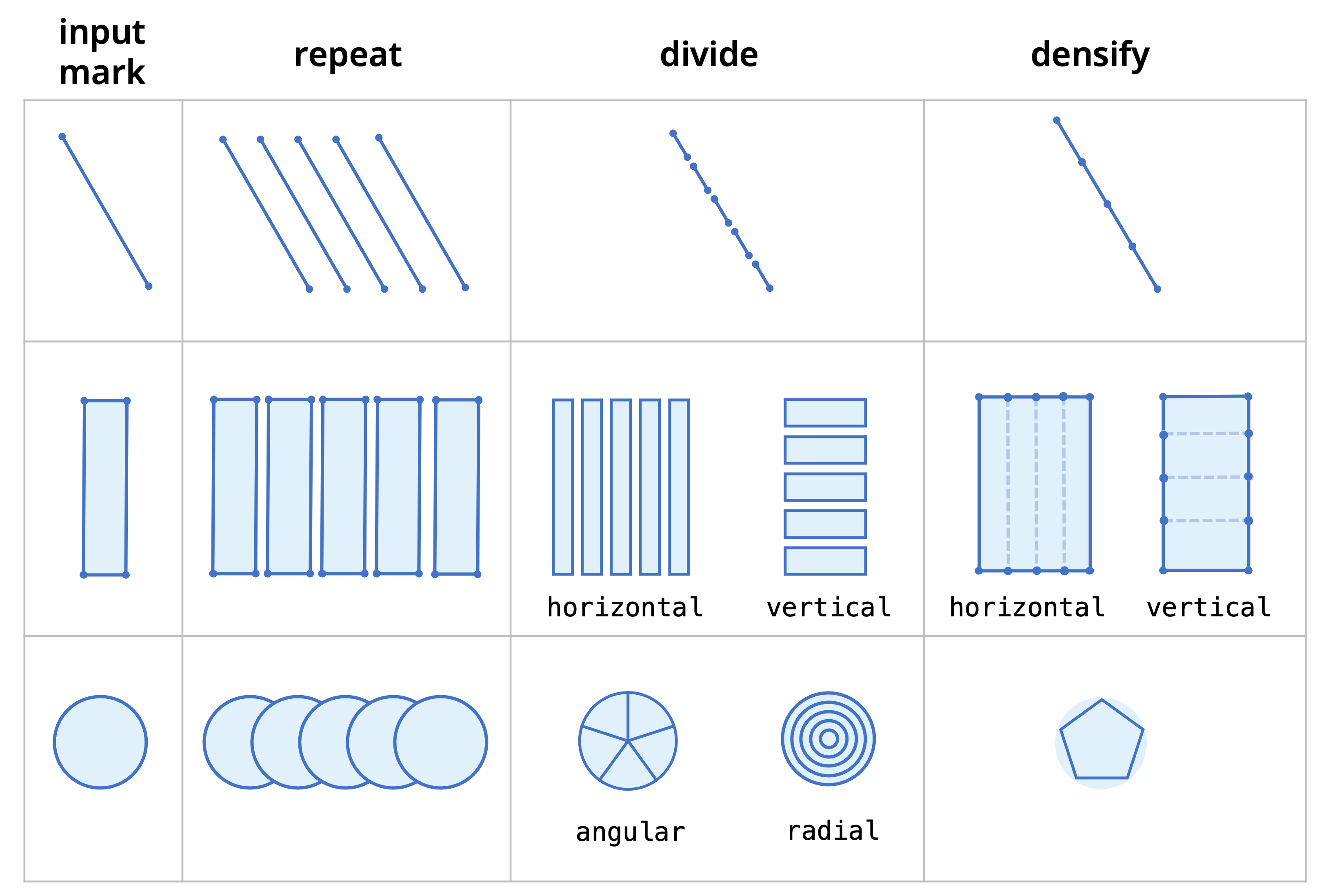} 
  \caption{Three operations for graphics-data join: repeat, divide, and \densify. Here each input mark is joined with five unique attribute values.\label{fig:join}}
  \vspace{-5mm}
\end{figure}

\bpstart{Repeat} The \rc{repeat} operation takes a mark, a glyph, or a collection as input and creates a collection of marks, glyphs, or collections (\cref{fig:join}). This operation accepts a dataset and an ordinal or nominal attribute name in the dataset as its parameters. \noindent{For} example, in \cref{fig:dsb_procedure}b, we repeat a rectangle by the \rc{age} attribute. Since there are four unique \rc{age} values in the data (below 30, 30 - 50, 50 - 70, above 70), we get a collection of four rectangles (\cref{fig:dsb_procedure}b). The data scope of the first rectangle consists of all the rows with ``below 30'' as the \rc{age} value, and so on. 

\noindent{Formally,} \noindent{when} the dataset is a table, the output of a \rc{repeat} operation is a collection $C = \{e_{1}, e_{2}, ..., e_{n}\}$, where each member visual element $e_{i}$ has the same type \revision{as} the input element. 
The number of elements in C is equal to the number of unique values in the specified parameter attribute $\alpha$. 
For each unique value $v_{i}$ in 
$\alpha$, there exists a visual element $e_{i}$ in $C$ such that $e_{i}.data[\alpha] = v_{i}$.

\getRule{repeat}{table, attr}{mark}{mark~collection}

\getRule{repeat}{table, attr}{glyph}{glyph ~collection}

\getRule{repeat}{table, attr}{collection}{nested ~collection}






\noindent{When} the dataset is a network, let $nd$ denote a visual element that represents an item, and $lk$ denote a mark (usually a line, a path, an arc, or a band) that represents a link: 

\getRule{repeat}{network, attr}{[nd, lk]}{[nd~collection, lk~collection]}


\vspace{2mm}
\bpstart{Divide} The \rc{divide} operation splits a mark into a collection of smaller marks. This operation accepts a data table, and an ordinal or nominal attribute name in the table as its parameters. An additional orientation parameter is needed when the input mark type is rectangle or circle. The mark type in the output collection depends on the input mark type and the orientation. 

\getRule{divide}{table, attr, orientation}{rect}{rect~collection}

\getRule{divide}{table, attr, orientation}{circle}{pie~collection~|~ring~collection}

\getRule{divide}{table, attr}{line}{line~collection}

\getRule{divide}{table, attr}{pie}{arc~collection}

\getRule{divide}{table, attr}{ring}{arc~collection}

\noindent{For} example, applying \rc{divide} on a circle along the \textit{angular} orientation produces a collection of pies, and along the \textit{radial} orientation produces a collection of rings (\cref{fig:join}). The graphics-data joining mechanism is similar to that of the \rc{repeat} operation: tuples sharing the same \rc{attribute} value are assigned as the data scope of each mark in the output collection.

\vspace{2mm}
\bpstart{Densify} \revision{In a line chart, area chart, or radar chart, we need to assign data scopes to the vertices of the geometric marks. MSC introduces the \rc{\densify} operation to facilitate the generation of vertices and the computation of data scopes.  The \rc{\densify} operation transforms a rectangle into an area mark, a circle into a polygon, and a line into a polyline (\cref{fig:join}) by adding vertices along the border of the input mark, assigning a data scope to each vertex, and replacing curve segments with line segments.} This operation accepts a data table and an ordinal or nominal attribute name in the table as its parameters. An additional orientation
parameter is needed when the input mark is a rectangle: vertices are added along parallel borders based on the specified orientation. 

\getRule{\densify}{table, attr}{line}{polyline}

\getRule{\densify}{table, attr}{circle}{polygon}

\getRule{\densify}{table, attr, orientation}{rect}{area}

\noindent{T}he graphics-data joining mechanism is similar to that of the \rc{repeat} operation and the \rc{divide} operation. Each vertex represents a unique value in the attribute. 




\vspace{2mm}

\bpstart{Classify} The \rc{classify} operation groups visual elements inside a collection into multiple collections. This operation accepts an ordinal or nominal attribute name as its parameter. 

\getRule{classify}{attr}{collection}{nested~collection}

\begin{figure}[h]
  \centering
  \vspace{-3mm}
  \includegraphics[width=0.9\linewidth]{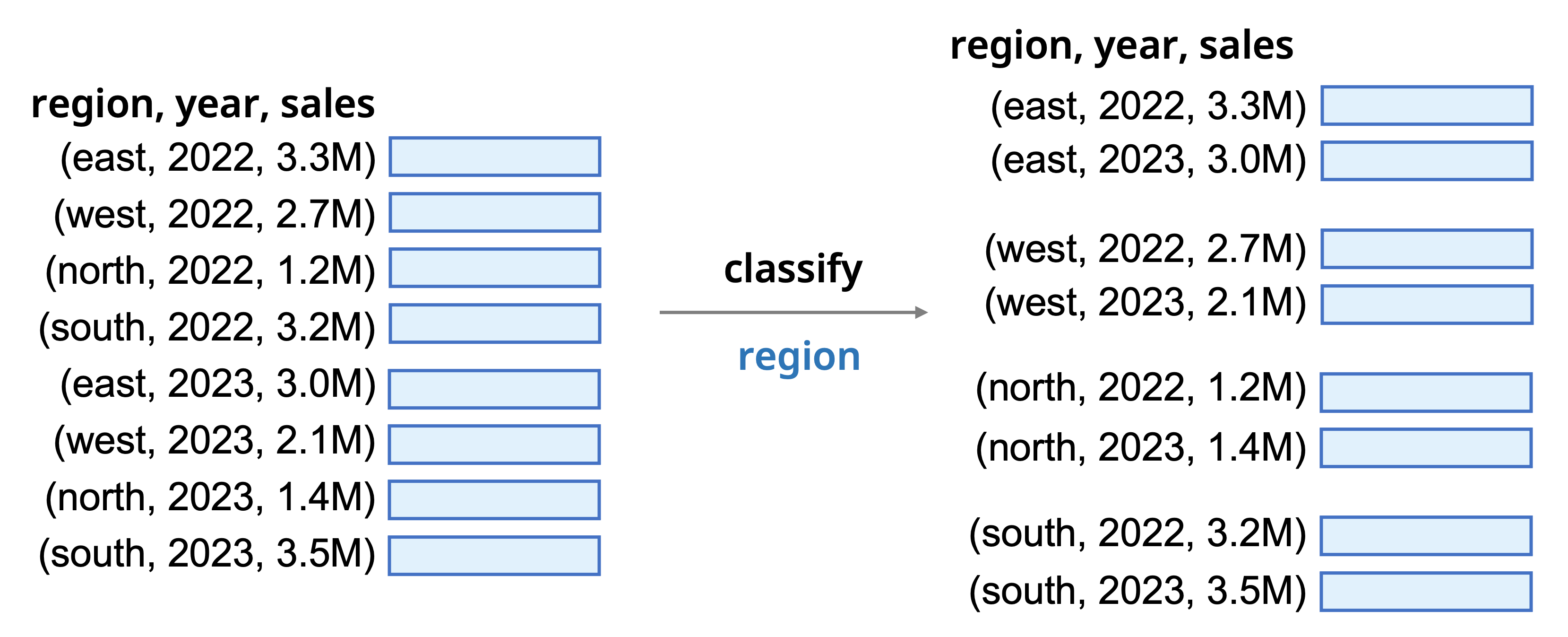} 
    \vspace{-3mm}
  \caption{Classifying a collection resulting in a nested collection. The text on the left of each mark denotes its data scope.  \label{fig:classify}}
\end{figure}

In \cref{fig:classify}, we classify a collection of eight rectangle marks by the \rc{region} attribute, resulting in a nested collection consisting of four member collections, each of which consists of two rectangle marks.

\noindent{Formally}, given an input collection 
 $C = \{e_{1}, e_{2}, ..., e_{n}\}$, classifying its members by an ordinal or nominal attribute $\alpha$ results in a nested collection, where the members of the input collection $C$ are now collections: $C = \{c_{1}, c_{2}, ..., c_{k}\}$. For each unique value $v_{i}$ in the specified parameter attribute $\alpha$, there exists a visual element $c_{i}$ in $C$ such that $c_{i}.data[\alpha] = v_{i}$. Each $c_{i}$ contains some of the original members (i.e., $e_{1}, e_{2}, ..., e_{n}$) such that for every member $e_{j}$ in $c_{i}$, $c_{i}.data[\alpha] = e_{j}.data[\alpha]$. 

\vspace{2mm}

\bpstart{Repopulate} \revision{A scene can be treated as a template, where the visualization design is applied to a new dataset.} The \rc{repopulate} operation \revision{enables such functionalities by re-generating} visual elements inside an existing collection based on a new attribute in a new dataset. This operation accepts \revision{pairs of ordinal/nominal attribute names (one from the new dataset and the other from the existing dataset) as its parameters.}

\getRule{repopulate}{\{(attr_{new}, attr_{current})\}, table_{new}}{collection}{collection'} 

\begin{figure}[ht]
  \centering
  \vspace{-3mm}
  \includegraphics[width=0.95\linewidth]{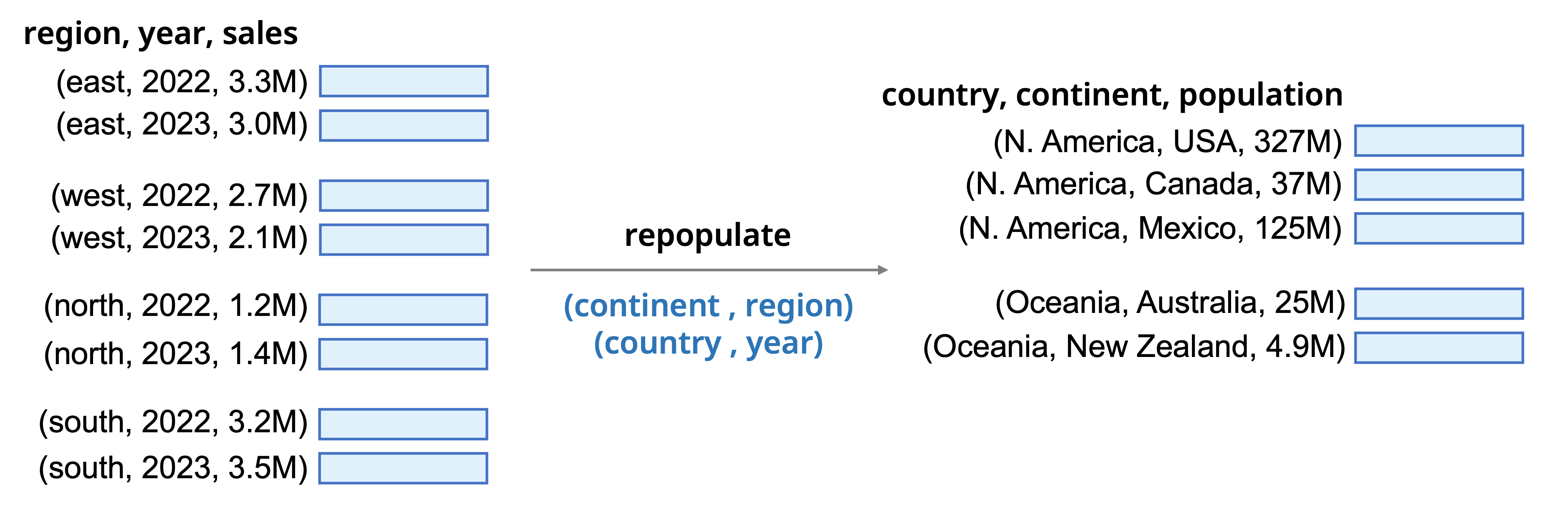} 
    \vspace{-3mm}
  \caption{Repopulating a nested collection where the number of member marks in the result changes based on the new dataset. \label{fig:repopulate}}
\end{figure}

\noindent\Cref{fig:repopulate} \revision{shows how to change the underlying dataset of a grouped chart from a table about regional sales to a new table about countries' populations using the \rc{repopulate} operation.} 
The text on the left of each mark denotes its data scope.  

\noindent{Formally}, given an input collection 
 $C = \{e_{1}, e_{2}, ..., e_{n}\}$, after applying the \rc{repopulate} operation by an attribute $\alpha$ from dataset $T$, we have a new collection $C' = \{e'_{1}, e'_{2}, ..., e'_{m}\}$. Every element $e'_{j}$ in $C'$ has the same type as any member $e_{i}$ in the input collection $C$, and the number of elements in $C'$ is equal to the number of unique values in $\alpha$. For each unique value $v_{i}$ in $\alpha$, there exists a visual element $e'_{i}$ in $C'$ such that $e'_{i}.data[\alpha] = v_{i}$.

\begin{figure}[ht]
 \vspace{-3mm}
  \centering
  \includegraphics[width=0.9\linewidth]{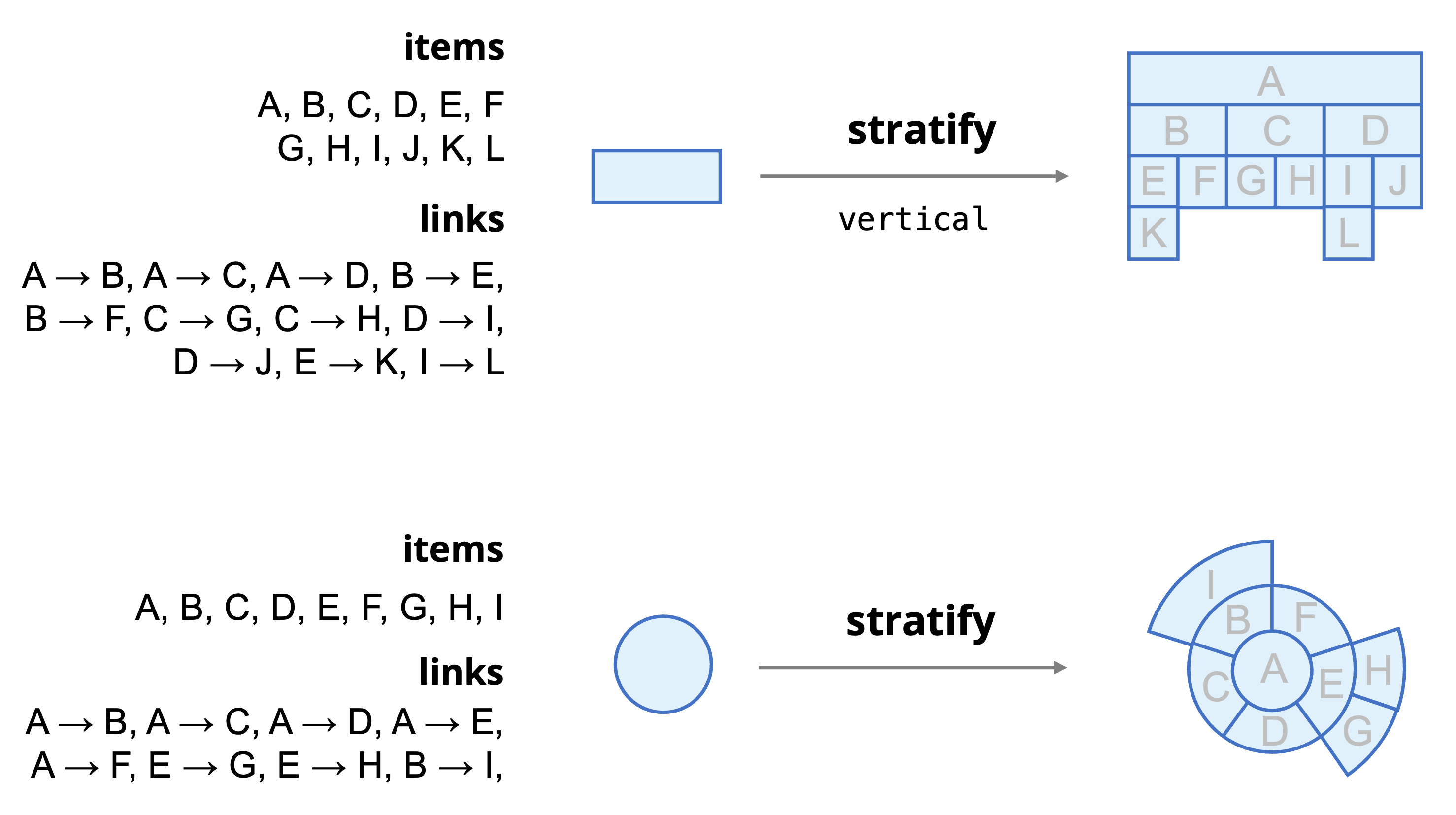} 
  \caption{Applying the stratify operation to a rectangle mark or a circle mark for a tree dataset.\label{fig:stratify}}
  \vspace{-3mm}
\end{figure}

\bpstart{Stratify} The \rc{stratify} operation takes a mark as
input and generates a collection of marks to be used as a layered representation of hierarchical data. This operation accepts a tree dataset (a special type of network dataset) and an ordinal or nominal attribute in the
dataset as its parameters. The layering of generated marks is determined by the links connecting the items in the tree dataset. When the input mark is a rectangle, an \rc{orientation} parameter is needed to specify the direction of layering. \revision{\Cref{fig:stratify} shows that stratifying a rectangle and a circle results in an icicle plot \cite{kruskal_icicle_1983} and a sunburst chart \cite{stasko_focus_2000} respectively.}

\getRule{stratify}{tree, attr, orientation}{rect}{rect~collection} 

\getRule{stratify}{tree, attr}{circle}{arc~ collection} 

\subsection{Modificative Operations: Specify Channel Values}
\label{sec:channel_operations}
\bpstart{Apply/Remove Visual Encoding} Mapping attribute values to visual channel properties is a fundamental task in data visualization. For instance, the width of each rectangle mark in \revision{\cref{fig:dsb_procedure}f} encodes the \rc{pct} value in the mark's data scope. Specifying such an operation in MSC requires the concept of \textit{peers}. The peers of a visual element $e$, denoted by $P_{e}$, is a set of all the visual elements generated together with $e$ through the \rc{repeat}, \rc{divide}, \rc{densify}, \rc{classify}, \rc{repopulate}, and \rc{stratify} operations in \cref{sec:join_operations}. For instance, all the four rectangle marks generated by \rc{repeat} in \revision{\cref{fig:dsb_procedure}b} are peers of each other; all the sixteen rectangle marks generated by \rc{divide} in \revision{\cref{fig:dsb_procedure}c} are peers of each other; all the vertices generated by a \rc{densify} operation are peers of each other. Binding a data attribute $\alpha$ with a channel of a visual element $e$ based on a scale $\lambda$ is expressed as:

\modificative{\forall e_{i} \in P_{e}, e_{i}.channel = \lambda(e_{i}.data[\alpha])}


\noindent{In} MSC, the element $e$ can be a vertex, a segment, a mark, a glyph, or a collection. This operation applies to both the initial mapping declaration as well as the update of channel values when the underlying data values for an existing mapping change. 


\bpstart{Customize Scale} A scale directly determines \textit{how} data attribute values are transformed into visual channel values. Assuming the data attribute values are fixed, we can specify the desired channel values by customizing the \textit{type} \revision{(\eg linear, power, log)}, \textit{domain} \revision{(e.g., start from zero)}, and \textit{range} \revision{(e.g., range extent for size, scheme for color)} of scales.

\bpstart{Set Channel Values Directly} In the absence of any data attribute bound to a visual element's channel, the value of that channel can be freely changed. It is a common practice to modify the channel values for all the marks in a data visualization. For instance,  the height of each rectangle mark in Figure 2(f) can be set to a constant $k$. Such an operation in MSC is expressed as:

\modificative{\forall e_{i} \in P_{e}, e_{i}.channel = k}

\noindent{In} certain cases, we may want to set the channel value for a specific visual element (e.g., for annotation purposes, to translate a visual element in design exploration) without affecting its peers, such an operation is expressed as: \modificative{e.channel = k}

\subsection{Modificative Operations: Apply Layouts \& Constraints} 
\label{sec:spatial_operations}

\bpstart{Apply Layout}
In MSC, layouts are properties of visual element groups, and can be directly changed. A layout type and a set of associated parameters are required for a group $G$:

\modificative{G.layout = \{type, parameters\}}

Similar to the operation to set channel values, we can set the layout of multiple groups that are peers to each other:

\modificative{\forall G_{i} \in P_{G}, G_{i}.layout = \{type, parameters\}}





\bpstart{Update Layout Parameter} A parameter $\theta$ of an existing layout can be dynamically set to a new value $v$. For example, we can change the orientation of a stack layout from horizontal to vertical, or change the number of rows in a grid layout:

\modificative{G.layout.\theta = v}

\bpstart{Apply Relational Constraint} 
This operation applies to both initializing a new relational constraint as well as updating the channel values to maintain an existing constraint. Relational constraints in MSC are expressed using arithmetic operators. For example, to align the left sides of two marks' bounding boxes:

\modificative{m_{i}.bbox.left = m_{j}.bbox.left}

\noindent{T}o affix the bottom left corner of a mark's bounding box with reference to the top right corner of another mark's bounding box:

\modificative{m_{i}.bbox.left = m_{j}.bbox.right + dx}

\modificative{m_{i}.bbox.bottom = m_{j}.bbox.top + dy}


In data visualization, relational constraints are usually applied to multiple pairs of visual elements. For instance, to attach a text label to the center of every rectangle mark in \cref{fig:dsb}a:

\modificative{\forall rect_{i} \in P_{rect},~\forall text_{i} \in P_{text},}

\modificative{rect_{i}.x = text_{i}.x~and~rect_{i}.y = text_{i}.y,}

\modificative{where~rect_{i}.data = text_{i}.data}








\subsection{Modificative Operations: Configure View}
\label{sec:camera_operations}

Finally, view configuration of a scene can be expressed by setting its properties, for example:

\modificative{scene.view.rotation = 90^{\circ}}

\modificative{scene.view.zoom~*= 1.2}

\subsection{Chaining Operations}
A chart construction or manipulation task often requires chaining multiple operations described above. Here we provide two examples where the operations can be combined to achieve higher-level tasks. 

\bpstart{Chaining Generative Operations}
\noindent{C}oncatenating these graphics-data join operations will produce nested structures. For instance, in \cref{fig:dsb}, we perform \rc{repeat} first followed by \rc{divide} to obtain a stacked bar structure; applying the \rc{densify} operation on a rectangle followed by the \rc{repeat} operation yields a structure for small multiples of area charts. 

\begin{figure}[ht]
 \vspace{-3mm}
  \centering
  \includegraphics[width=\linewidth]{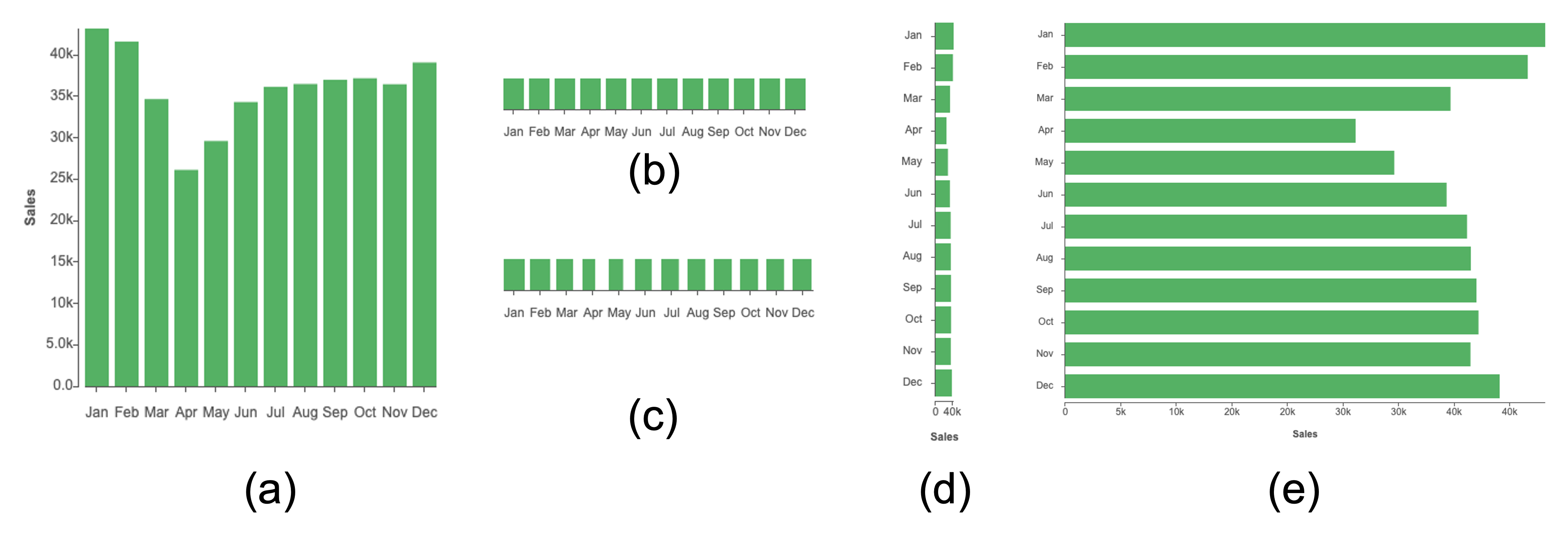} 
   \vspace{-5mm}
  \caption{Chaining modificative operations to change the orientation of a bar chart. \label{fig:chain}}
\end{figure}

\bpstart{Chaining Modificative Operations}
Modifying the design of a chart may involve more than one operation. 
For instance, we may want to change the vertical bar chart in \cref{fig:chain}a into a horizontal bar chart in \cref{fig:chain}e, which can be accomplished through the following sequence of operations: \rc{Set~Bar~Height~to~Constant} (\cref{fig:chain}b), \rc{Bind~Attribute~to~Width} (\cref{fig:chain}c), \rc{Update~Grid~Layout~Parameter} (\cref{fig:chain}d), and \rc{Customize~Width~Scale} (\cref{fig:chain}e).

\section{Implementation: Mascot.js}
We implemented a JavaScript library, Mascot.js, based on the theoretical descriptions on the components (\cref{sec:components}) and operations (\cref{sec:procedure}) in MSC. A catalog of visualization examples created using Mascot.js is available at \url{https://mascot-vis.github.io}, which showcases more than 60 unique data visualization designs, covering commonly used chart types as well as more esoteric designs like cycle plot \cite{robbins_introduction_2008} and treemap bar chart.


\bpstart{API Design} 
Mascot.js provides global functions to import datasets,
initialize scene objects,
and create layouts,
The generative operations are invoked as scene methods, operating within the context of that scene.
Modificative operations applied to a single element are invoked through the element's setter method, 
whereas modificative operations applied to multiple elements or an element's peers are invoked through the appropriate methods in the scene object. The details on the API design are available in our short paper on a prior implementation of MSC called Atlas.js \cite{liu_atlas_2021}.

\bpstart{Smart Defaults} Mascot.js chooses smart default values if certain properties or components are not specified. For example, when creating new marks, default values for various channels are applied depending on the mark type. When a \rc{divide} is performed, Mascot.js will apply a default layout to the resulting collection depending on the mark type. For instance, dividing a rectangle will also apply a stack layout to the generated smaller rectangles.

\bpstart{Creating Auxiliary Elements} Mascot.js provides convenience methods to create axes and legends. Users only need to specify the channel, attribute, and preferred properties such as position and visibility. 

\bpstart{Rendering} Unlike libraries such as D3 \cite{bostock_d3_2011} where visualization objects are represented as Document Object Model (DOM) nodes, Mascot.js represents visualization scenes as JavaScript objects and decouples scene abstractions
from rendering. All the objects in a scene can be printed to the console and inspected without rendering. To display these objects, we need to create renderers. 
Mascot.js supports two types of renderers: SVG and WebGL.

\bpstart{Scene Serialization and Reconstruction} Scenes created using Mascot.js are JavaScript objects and can be serialized as JSON files. Mascot.js provides the functionality to ingest such JSON files and reconstruct the scene objects so that they can continue to be modified using MSC operations. The JSON files can also serve as templates: new visualizations can be created based on new datasets by applying a series of \rc{repopulate} and \rc{apply~visual~encoding} operations on the templates.

\section{Applications}
\label{sec:applications}
In this section, we examine the potential of MSC as a foundation for three applications through case studies: interactive authoring of static representations, deconstructing and reusing existing charts, and augmenting static visualizations with animated behaviors. 

\subsection{Interactive Chart Authoring}
Multiple research projects seek to design and build interactive authoring tools for users to construct static data visualizations in a WYSIWYG (What You See is What You Get) interface \cite{satyanarayan_lyra_2014,ren_charticulator_2018,liu_data_2018,kim_data-driven_2016,ren_ivisdesigner_2014,tsandilas_structgraphics_2020}. 
To demonstrate the viability of MSC in powering real-world authoring applications, we implemented a new version of Data Illustrator \cite{liu_data_2018} based on Mascot.js, which is available at \url{https://data-illustrateur.github.io/}. The benefits of using MSC as the core computational representation are twofold. First, thanks to a more complete theoretical formulation offered by MSC on data visualization components and operations, the new version of Data Illustrator is more expressive, supporting a range of data visualization designs that were not possible in the original version, including but not limited to rose chart, area chart, icicle plot, sunburst chart, and network diagrams. Second, MSC abstracts away the low-level complexities in scene manipulation and reduces development time and effort. The original Data Illustrator was built on Paper.js \cite{lehni_paperjs_nodate}, which provides general scene graph abstraction support without visualization-specific components and operations. Compared to this original implementation involving 22632 lines of code in JavaScript, the new Data Illustrator implementation based on Mascot.js has only 5519 lines of code---more than four times reduction.

\subsection{Chart Deconstruction and Reuse}
Since constructing visualizations from scratch can still require significant knowledge and skills in using a particular authoring tool, researchers have also investigated methods and systems that automatically analyze existing visualizations and transform them into reusable templates \cite{chen_mystique_2023,harper_converting_2017,harper_deconstructing_2014,cui_mixed-initiative_2022,chen_towards_2020,poco_reverse-engineering_2017,poco_extracting_2018}.

We investigated the potential of MSC as a computational representation for visualization reuse in the Mystique system \cite{chen_mystique_2023}. Mystique takes existing rectangle-based visualizations in the SVG format as input, and extracts scene components including reference elements (axis \& legend), groups (collection \& glyph), algorithmic layouts (grid, stack, packing), and relational constraints (alignment \& affixation). The extracted components can then be manipulated and infused with a new dataset to produce a new scene through operations such as \rc{repopulate}, \rc{apply~visual~encoding}, and \rc{apply~layout~\&~constraint}. Mystique provides a wizard interface to guide users in specifying the mappings between the components and the new dataset, and produce immediate visual feedback by updating the input example with the incrementally specified mappings. The benefits of using MSC as the computational representation for visualization deconstruction and reuse include the following: 1) we can directly establish correspondence between SVG elements and scene components, without having to go through an intermediate formal language, 2) the direct representation of scenes allows us to dive deep into complex layout structures such as nesting and small multiples, and 3) the intermediate scene results generated in the reuse process do not have to conform to the semantics and syntax of a formal language, allowing users focus on one mapping at a time. 




\subsection{Animating Static Visualizations}

Besides authoring and reusing static visualizations, there is growing interest in research works to augment static visualizations with interactive and animated behaviors \cite{thompson_data_2021,wang_animated_2021,snyder_divi_2024,ge_canis_2020,ge_cast_2021,zong_lyra_2020}. To enable such augmentation, there exists two approaches to ingest the static visualizations. The first approach requires that the static visualizations must be created by a particular tool. For example, Data Animator \cite{thompson_data_2021} adds animated behaviors to static visualizations created using Data Illustrator \cite{liu_data_2018}, and Lyra 2 \cite{zong_lyra_2020} turns static visualizations composed using Vega-Lite \cite{satyanarayan_vega-lite_2016} into interactive ones. Such assumptions ensure that the semantic components are readily usable for behavior specification.  Alternatively, we can broaden the scope to support input static visualizations in a specific format generated by multiple tools, but the visualizations need to be annotated with semantic component labels.
For instance, CAST \cite{ge_cast_2021} is an animation authoring tool and works with SVG charts, provided that these charts are annotated with labels in a format called dSVG (data-enriched SVG) \cite{ge_canis_2020}. The dSVG format used by CAST can be considered an attempt to formulate a computational representation, however, it is not well documented and is a one-time solution for this specific task of animation authoring.

We are interested in understanding how well a more generic computational representation like MSC supports animation behavior specification. In this case study, we focus on two existing systems: Data Animator and CAST. As mentioned earlier, Data Animator requires static visualizations to be created using Data Illustrator, and MSC's component formulation is a more complete account and encompasses the component model of Data Illustrator, it is thus trivial to convert visualizations expressed using MSC into the Data Illustrator format. CAST, on the other hand, uses dSVG as the underlying computational representation.  
Unlike MSC, dSVG does not explicitly describe the hierarchical relationships between scene elements, and adds three properties ``id'', ``class'', and ``datum'' for each mark. Based on the grouping and data scope components in MSC, we were able to develop scripts that convert data visualization scenes created using Mascot.js into the dSVG format, and verified that the resulting scenes in the dSVG format worked well with the CAST system. The supplemental materials contain the conversion scripts as well as video demos demonstrating the specification of animation behaviors. This exercise validates MSC as a computational representation for augmenting static visualizations, demonstrating its compatibility with current animation authoring tools and its potential as a foundation to develop similar systems in the future.



\section{Comparison with Related Frameworks}

\revision{As discussed in the introduction, the goal of proposing MSC as a new computational representation for data visualization scenes is to support various applications for constructing, decomposing, and augmenting visualizations. While Mascot.js can be considered yet another data visualization library, its usability and learnability as a domain-specific language are not our primary concern. 
Instead, we are interested in understanding the strengths and weaknesses of MSC as a theoretical framework in enabling various applications. To evaluate MSC, we first present an overall comparison with nine visualization frameworks in terms of component and operation abstractions. We then focus on more in-depth comparisons with three classes of related work: 
DOM manipulation libraries (D3), declarative  languages (e.g., Vega/Vega-Lite), and algebraic frameworks (e.g., VizQL).} 

\begin{table*}[!htb]\fontfamily{ptm}\selectfont\small
\centering
\caption{A comparison of the primitives (i.e., components and operations) defined in various data visualization frameworks. A \greenCheck~means the primitive is included in the framework, but the paper does not formally name it. If a primitive is named, we include the name used in the paper in the corresponding cell. A 
\cross~means the primitive is not included in the paper.
}

\resizebox{\textwidth}{!}{\small%
\begin{tabular}{|l|l|l|l|l|l|l|l|l|l|l|l|}
\arrayrulecolor{lightgray}
\hline
\rowcolor[gray]{.95} & \textbf{MSC} & \textbf{Protovis} \cite{bostock_protovis_2009} & \textbf{D3} \cite{bostock_d3_2011} & \textbf{Vega} \cite{noauthor_vega_2021} & \textbf{\begin{tabular}[c]{@{}l@{}}Data Illustrator\end{tabular}} \cite{liu_data_2018} & \textbf{Charticulator} \cite{ren_charticulator_2018} & \textbf{Bluefish} \cite{pollock_bluefish_2023} & \textbf{Chartreuse} \cite{cui_mixed-initiative_2022} & 
\textbf{AutoTimeline} \cite{chen_towards_2020}
 & \textbf{InfoMotion} \cite{wang_animated_2021}   \\
\hline
 \multirow{12}{*}{\rotatebox[origin=c]{90}{Components}} & mark & mark & mark & mark & mark & mark & element & \begin{tabular}[c]{@{}l@{}}visual element\end{tabular} & mark  & \begin{tabular}[c]{@{}l@{}}visual element\end{tabular}   \\
\cline{2-11} 
 & data scope & \cross & data & \cross & data scope & \greenCheck & \cross  & \cross & \cross  & \cross  \\
\cline{2-11} 
 & glyph & \cross & \cross & \cross & group & glyph & group & unit & \cross & \begin{tabular}[c]{@{}l@{}}repeating unit\end{tabular}   \\
\cline{2-11} 
 & collection & panel & \cross & \cross & collection & \greenCheck & group & \cross & layout  & \begin{tabular}[c]{@{}l@{}}group\end{tabular} \\
\cline{2-11} 
 & composite & panel & \cross & \cross & \begin{tabular}[c]{@{}l@{}}\cross\end{tabular} & \cross & \begin{tabular}[c]{@{}l@{}}\cross\end{tabular} & \cross & \cross & \begin{tabular}[c]{@{}l@{}}\cross\end{tabular}  \\
\cline{2-11} 
 & \begin{tabular}[c]{@{}l@{}}reference element\end{tabular} & rule & axis, legend & axis, legend & \begin{tabular}[c]{@{}l@{}}axis, legend\end{tabular} & \begin{tabular}[c]{@{}l@{}}axis, legend
 \end{tabular} & \cross  & \cross & \cross  & \cross  \\
\cline{2-11} 
 & annotation & \cross & \cross & \cross & \cross & \greenCheck & \begin{tabular}[c]{@{}l@{}}connectedness,\\common region\end{tabular} & \begin{tabular}[c]{@{}l@{}}embellishment\end{tabular} & annotation & connectors  \\
\cline{2-11} 
 & \begin{tabular}[c]{@{}l@{}}visual encoding\end{tabular} & \greenCheck & \cross & \greenCheck & \greenCheck & visual encoding & \greenCheck  & \begin{tabular}[c]{@{}l@{}}visual element\end{tabular} & \begin{tabular}[c]{@{}l@{}}visual encoding\end{tabular} & \cross   \\
\cline{2-11} 
 & scale & scale & scale & scale & \greenCheck & scale & \cross & \cross & scale & \cross   \\
\cline{2-11} 
 & \begin{tabular}[c]{@{}l@{}}algorithmic\\layouts\end{tabular} & layouts & layouts & layouts & layouts & \begin{tabular}[c]{@{}l@{}}constraint-based\\layout\end{tabular} & \cross & \begin{tabular}[c]{@{}l@{}}unit\\layout\end{tabular} & representation  & \begin{tabular}[c]{@{}l@{}}unit\\layout\end{tabular} \\
\cline{2-11} 
 & \begin{tabular}[c]{@{}l@{}}relational constraints\end{tabular} & anchor & \cross & \cross & \greenCheck & \greenCheck & \begin{tabular}[c]{@{}l@{}}perceptual relation\end{tabular}  & unit layout & \cross & \begin{tabular}[c]{@{}l@{}}unit layout\end{tabular}  \\
\cline{2-11} 
 & view configuration & \cross & \cross & \greenCheck & \cross & \cross & \cross & \cross & \cross & \cross  \\[2.5pt]
\Xhline{2\arrayrulewidth}
\multirow{13}{*}{\rotatebox[origin=c]{90}{Operations}} & repeat \rule{0pt}{3ex} & \cross & join & \cross & \greenCheck & \greenCheck & \greenCheck & \cross & \cross & \cross   \\
\cline{2-11} 
 & divide & \cross & \cross & \cross & \greenCheck & \cross & \cross  & \cross & \cross & \cross  \\
\cline{2-11} 
 & densify & \cross & \cross & \cross & \cross & \cross & \cross & \cross & \cross & \cross \\
\cline{2-11} 
 & classify &  \cross & \cross & \cross & \cross & \cross & \cross & \cross & \cross & \cross  \\
\cline{2-11} 
 & repopulate & \cross  & \cross & \greenCheck & \cross & \greenCheck & \cross & repeat & \greenCheck & \cross  \\
\cline{2-11} 
 & stratify & \cross & \cross & \greenCheck & \cross & \cross & \cross & \cross & \cross & \cross  \\
\cline{2-11} 
 & \begin{tabular}[c]{@{}l@{}}apply visual encoding\end{tabular} & \greenCheck & \greenCheck & \greenCheck & \greenCheck & \greenCheck & \greenCheck & morph, recolor & \greenCheck & \cross  \\
\cline{2-11} 
 & customize scale & \greenCheck & \greenCheck & \greenCheck & \greenCheck & \greenCheck & \cross  & \cross & \cross & \cross  \\
\cline{2-11} 
 & \begin{tabular}[c]{@{}l@{}}set channel values\end{tabular} & \greenCheck & \greenCheck & \greenCheck & \greenCheck & \greenCheck & \greenCheck & fix & \cross & \cross   \\
\cline{2-11} 
 & apply layout & \greenCheck & \greenCheck & \greenCheck & \greenCheck & \greenCheck & \greenCheck & partition & \greenCheck  & \cross  \\
\cline{2-11} 
 & \begin{tabular}[c]{@{}l@{}}update layout\\parameters\end{tabular} & \greenCheck & \greenCheck & \greenCheck & \greenCheck & \greenCheck & \greenCheck & \cross & \cross & \cross  \\
\cline{2-11} 
 & \begin{tabular}[c]{@{}l@{}}apply relational\\constraint\end{tabular} & \greenCheck & \cross & \cross & \greenCheck & \greenCheck & \greenCheck & move & \cross  & \cross   \\
\cline{2-11} 
 & \begin{tabular}[c]{@{}l@{}}configure view\end{tabular} & \cross & \cross & \greenCheck & \cross & \cross & \cross & \cross & \cross & \cross  \\
\hline
\end{tabular}
}

\label{table:comparison}
\end{table*}

\subsection{Component and Operation Abstractions}
\revision{Many visualization libraries and tools have proposed scene abstractions for various applications. Protovis \cite{bostock_protovis_2009} argues that designers should be able to think in terms of graphical marks, not abstract specifications to improve accessibility, and MSC shares the same stance. The components and operations in MSC are directly informed by recent visualization authoring frameworks and tools \cite{satyanarayan_critical_2019}. Specifically, the graphics-centric approach in MSC is consistent with the lazy data binding approach in Data Illustrator \cite{liu_data_2018}, and a few operations in MSC are derived from the repeat and partition operators in Data Illustrator \cite{liu_data_2018}; Charticulator's constraint-based layout approach \cite{ren_charticulator_2018} inspired the modificative operations in MSC.} 

\revision{To assess the completeness of abstraction, we compare MSC with nine visualization frameworks (\cref{table:comparison}), focusing on visual representations instead of interaction or animation. 
Among these, six were developed primarily for visualization creation: Protovis \cite{bostock_protovis_2009}, D3 \cite{bostock_d3_2011}, Vega \cite{noauthor_vega_2021}, Data Illustrator \cite{liu_data_2018}, Charticulator \cite{ren_charticulator_2018}, and  Bluefish \cite{pollock_bluefish_2023}; two were developed primarily for infographics reuse \cite{cui_mixed-initiative_2022,chen_towards_2020}; and one was developed for augmenting visualization with animation \cite{wang_animated_2021}.}
\revision{In general, MSC offers a more complete set of components and operations (\cref{table:comparison}). It is important to note that completeness is not equivalent to expressiveness: a missing component or operation in a framework does not mean that the corresponding visualization feature cannot be realized using the framework. For example, D3 provides very few abstractions but is highly expressive due to its reliance on the DOM as the scene representation. In addition, due to the different philosophies and approaches adopted by these frameworks, a direct one-to-one comparison is not always possible. For example, Vega \cite{noauthor_vega_2021} does not provide some generative operations because it is essentially a declarative specification language, not a scene representation. We elaborate on the differences between MSC and D3 \cite{bostock_d3_2011} in \cref{sec:d3} and MSC and Vega in \cref{sec:specs}.
}


\subsection{Data-Driven Documents (D3)}
\label{sec:d3}

\revision{The main goal of D3 \cite{bostock_d3_2011} is to support data-driven manipulation of documents represented using the DOM (document object model) standard. 
D3 provides very limited scene abstraction for data visualization, offering only five types of components (mark, data, axis/legend, scale, and layout). Everything else needs to be represented and manipulated as SVG elements through low-level methods.
Such an approach enables an unparalleled level of expressiveness compared to other visualization frameworks. However, neither the low-level JavaScript code nor the resultant SVG scene graph serve as effective computational representations for applications beyond scene assembly. 
For instance,
D3 provides no encapsulation for components like visual encodings or relational constraints:  elements' visual properties need to be 
computed using low-level code, and any subsequent manipulation of the SVG scene graph will not automatically update the element properties to preserve the encodings or constraints. The lack of higher-level abstraction makes it challenging to build graphical user interfaces on top of D3 to support interactive authoring from scratch; to date, no such tools have been created. Similarly, without high-level encapsulation of components and operations, efforts to deconstruct and reuse D3 visualizations mostly focus on visual styles \cite{harper_deconstructing_2014,harper_converting_2017}, and cannot be easily extended to structural features like layouts \cite{chen_mystique_2023}.}

\subsection{Declarative Specifications}
\label{sec:specs}
\revision{Compared to D3, visualization grammars such as ggplot2 \cite{wickham_ggplot2_2009} and Vega \cite{satyanarayan_reactive_2016,satyanarayan_vega-lite_2016} provide higher-level and more systematic abstractions.  In these approaches, the abstractions are in the form of declarative specifications that succinctly describe the data transformations and visual encodings in a visualization. Most of these declarative specifications are implementations of the Grammar of Graphics \cite{wilkinson_grammar_2005}, which defines grammatical components as ``rules for constructing graphs mathematically''. In contrast, MSC describes the semantic structures and operations from a graphics-centric perspective. For example, the Grammar of Graphics describes a pie chart as the result of mapping data categories to rectangles and performing a polar transformation \cite{wilkinson_grammar_2005}; under the MSC paradigm, we may describe the process as dividing a circle and assigning categories to each pie, which is more interpretable.} 



\revision{ 
Unlike MSC and D3, declarative specifications are not representations of scenes. They are descriptions of what the scenes should be like---as Wilkinson put it, ``\textit{A scene and its description are different}'' \cite{wilkinson_grammar_2005} (p.7). Specifications need to be compiled into scenes, which are typically represented and rendered using SVG. While grammars like Vega do have internal representations of scenes, they hide details such as the structural relationships between visual elements from users. Any modification to the scene must be done by changing the specification, and users have no means to select or modify scene components directly. We elaborate on the implications of this difference for two sample applications below:  
}

\begin{figure}[ht]
  \centering
  \includegraphics[width=0.95\linewidth]{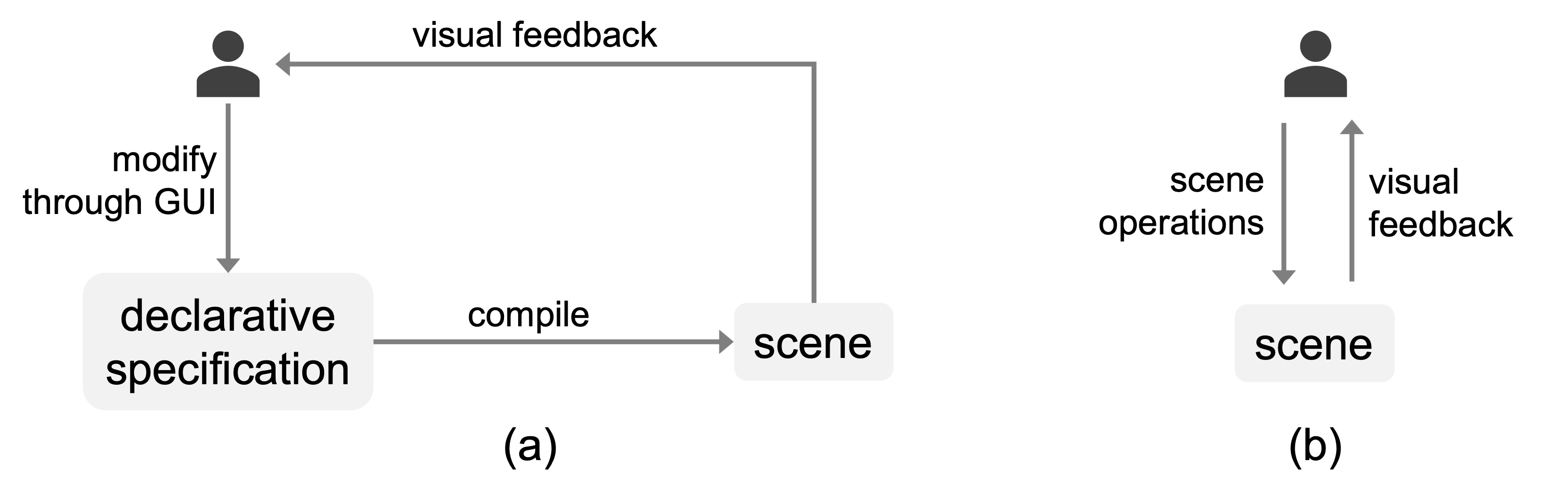} 
  \caption{Interactive authoring using (a) declarative specifications and (b) MSC as the underlying computational representations. \label{fig:authoring}}
\end{figure}

\noindent\revision{
\textbf{Interactive Authoring.} In Lyra \cite{satyanarayan_lyra_2014}, Vega serves as the underlying computational representation: user interaction in the GUI translates to changes in an underlying specification, which is dynamically compiled to a visualization scene (\cref{fig:authoring}a). In comparison, MSC provides a more direct experience in scene manipulation where the intermediate specifications are eliminated (\cref{fig:authoring}b). The MSC approach has several potential benefits: 1) during the authoring process, users can more freely manipulate the visualization scene, which may not conform to the semantics and syntax of a declarative specification language. Consequently, the authoring interface affords less restricted creative exploration and tinkering \cite{louridas_design_1999}, 2) the explicit conceptualization of semantic components such as collection and constraints afford direct manipulation interaction designs, and 3) there is no need to re-compile the entire specification every time a minor scene modification happens, which can lead to a more responsive and lower-latency experience.} 

\begin{figure}[ht]
  \centering
  \includegraphics[width=0.95\linewidth]{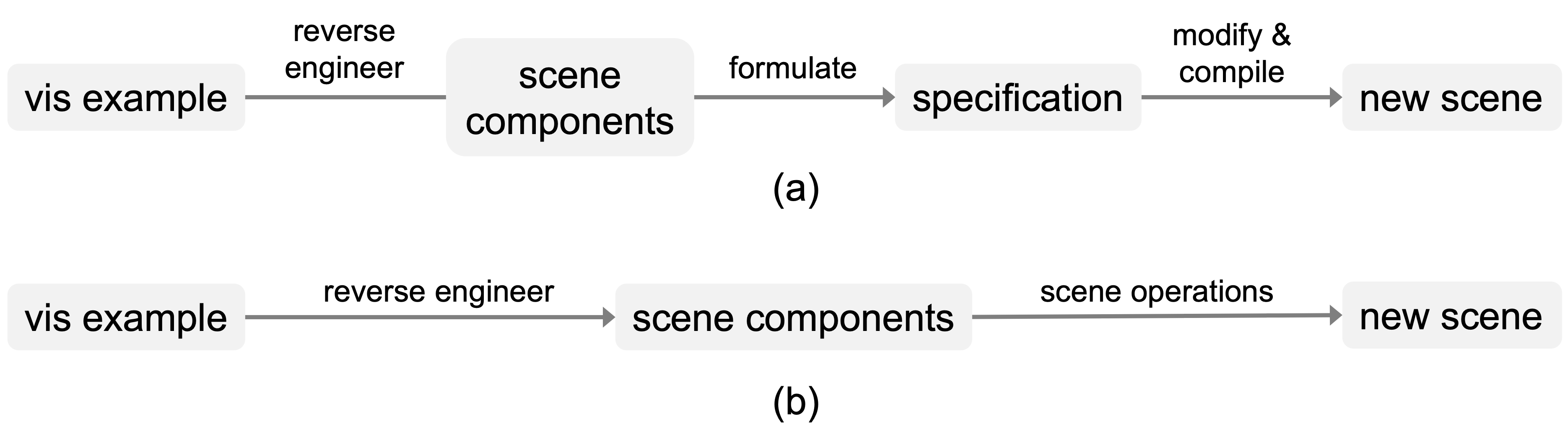} 
  \caption{Deconstructing and reusing existing visualizations using (a) declarative specifications and (b) MSC as the underlying computational representations. 
  \vspace{-4mm}
  \label{fig:reuse}
  }
\end{figure}

\noindent\revision{
\textbf{Visualization Deconstruction and Reuse.} A typical processing pipeline for this application involves two stages: 1) reverse engineer an existing data visualization to understand its structure, 2) generate a new visualization of a new dataset based on this understanding. The input visualizations are usually in a raster image or vector graphics format. The output of the reverse engineering stage and the input of the generation stage can be expressed either as declarative specifications (e.g., \cite{poco_reverse-engineering_2017}) or scene abstractions (e.g., \cite{cui_mixed-initiative_2022,chen_mystique_2023,chen_towards_2020}).} \revision{
In Mystique \cite{chen_mystique_2023}, we demonstrated how the reconstruction results for SVG charts with complex layout structures can be directly expressed using MSC, and the scene components can be dynamically updated through MSC operations to infuse with a new dataset (\revision{\cref{fig:reuse}b}). If declarative specifications were instead chosen as the computational representation, deconstruction and reuse would be more challenging: we would need to convert the deconstruction results into an intermediate specification (\revision{\cref{fig:reuse}a}), and dynamically generate partial specifications during the reuse process. Currently, no such solutions exist for charts with advanced layouts like small multiples and nested grouping}. 


\subsection{Algebraic Frameworks}
\revision{VizQL \cite{hanrahan_vizql_2006} is a formal language used by Tableau \cite{noauthor_tableau_nodate} to describe data visualizations. VizQL allows the specification of  table configurations (\ie rows and columns) as well as visual encodings within each pane. Specifically, table configurations are expressed using an algebra, where the operands are attributes (categorized into dimensions and measures) and the operators include cross, next, and concatenation \cite{stolte_polaris_2002}. 
}

\revision{The operations in MSC can be considered algebraic: the computation of data scopes when chaining the \rc{repeat}, \rc{divide}, and \rc{densify} operations is equivalent to the nest operator in VizQL; the \rc{classify} operation in MSC is similar to the \rc{group~by} operation in relational algebra; and the \rc{repeat} operation for network data is equivalent to the concatenation of the \rc{create~nodes} and \rc{create~connections} in the Ploceus framework \cite{liu_ploceus_2014,liu_network-based_2011}. However, in MSC operations, the primary operands are visual elements, and the attributes are secondary.
The generative operations in MSC describe how marks and groups are derived from input visual elements; the modificative operations in MSC describe the algebraic relationships between visual channel properties.}




\revision{The atttribute-centric algebra in VizQL represents a top-down approach to visualization generation: one starts from table configurations, then proceeds to mark choice and visual encodings in each table pane. This approach enables easy and systematic specifications of complex nested designs like small multiples. In contrast, the graphics-centric algebra in MSC is bottom-up: we start from marks and glyphs, and proceed to higher-level groupings. Additional effort is needed to apply the appropriate layouts to achieve table-based designs.}
\revision{In VisQL, since visual elements are not treated as first-class citizens, describing visualizations that do not fit neatly into a row-and-column metaphor can be challenging. For example, specifying glyph-based plots (e.g., box plot, dumbbell chart) involves non-trivial workarounds with custom attribute derivation and calculations. 
}


\section{Conclusion and Future Work}
To support data visualization applications such as reverse engineering and authoring, a computational representation is needed to abstract the components and their manipulations in a visualization scene. Manipulable Semantic Components (MSC) provide this by specifying a unified object model for the components in a visualization scene and a set of operations to generate and modify the components.


%
\revision{In future work, we plan to support both design-time and run-time interactions in MSC. The current focus of MSC is on design-time manipulation (e.g., construction and editing of visual representations); however, letting users specify run-time interactive behaviors (e.g., brushing, filtering) tailored for different scenarios (e.g., exploratory analysis, scrolly-telling) is essential. Designing a unified architecture for both kinds of interactions remains a challenge.  One potential solution is to model the relationships between the components as a dependency graph: whenever the value of a variable (e.g., a component's channel) is changed, the architecture should automatically propagate the change and update the values of other dependent variables. To capture the dependencies, we may formulate the relationships between components as constraints \cite{ren_charticulator_2018,vander_zanden_lessons_2001}: for example, the generative operations, layout operations, and encoding-related operations are one-way constraints, while operations that apply affixation and alignment can be described as multi-way constraints. Future research can focus on how to model and solve such constraint-based dependency graphs.}  

\revision{Another interesting future direction is to explore how to represent and learn design knowledge from visualizations expressed using MSC components. For example, Draco \cite{moritz_formalizing_2018} formalize and extract design knowledge from declarative specifications, and we are interested in understanding if Draco can be easily extended to work on scene representations such as MSC, and if novel methods can be developed to model design knowledge.  
}

\acknowledgments{%
Zhicheng Liu and Chen Chen were supported in part by NSF grant IIS-2239130.
}

\bibliographystyle{abbrv-doi-hyperref}

\bibliography{references}

\end{document}